\documentclass[aps, prmaterials, twocolumn, longbibliography, 10pt]{revtex4-2}

\usepackage[a4paper, total={7in, 10.3in}]{geometry}
\usepackage[utf8]{inputenc}
\usepackage[english]{babel}
\usepackage[font=small,labelfont=bf,  justification=raggedright,singlelinecheck=false]{caption} 
\usepackage{amsfonts, amsmath, amsthm, amssymb} 
\usepackage{braket}
\usepackage{graphicx}
\usepackage{footnote}
\usepackage{tabularx}
\graphicspath{{Figures/}} 
\usepackage[plainpages=false,pagebackref=false,pdftex]{hyperref}

\hypersetup{colorlinks=true,breaklinks=true,linkcolor=black,citecolor=black, urlcolor=black}
\usepackage[a-1b]{pdfx}

\begin{document}
    
\title{Revisiting ab-initio excited state forces from many-body Green's function formalism: approximations and benchmark}

\author{Rafael R. Del Grande}
\email{rdelgrande@ucmerced.edu}
\author{David A. Strubbe}
\email{dstrubbe@ucmerced.edu}
\affiliation{Department of Physics, University of California, Merced, California, USA}

\date{\today}

\begin{abstract}

Ab initio techniques for studying the optical and vibrational properties of materials are well-established, but only a few recent studies have focused on the interaction between excitons and atomic vibrations. In this paper, we revisit the excited state forces method, based on GW/BSE and DFPT calculations and provide a practical implementation and straightforward workflow. We fixed issues from Ismail-Beigi and Louie's implementation in \textcolor{blue}{Phys. Rev. Lett. 90, 076401 (2003)} and use an approximation for GW-level electron-phonon coefficients that improves our calculations accuracy. We explore its technical aspects, including convergence and the quality of approximations used for CO molecule, LiF and monolayer MoS$_2$. We successfully apply this method to investigate diverse kinds of self-trapped excitons in LiF, including polaronic excitons, discuss excited state relaxation strategies and project excited state forces in phonon displacement to explore exciton-phonon interactions. Our results provide the tools to study exciton-phonon related phenomena in molecules, low-dimensional materials and solids, including coherent phonon generation, self-trapped exciton and Stokes shift.

\end{abstract}

\maketitle


\section{Introduction}

Light-induced changes in materials play an important role in photovoltaic degradation, Stokes shifts, exciton transport \cite{Cohen2024, Chan2023}, exciton localization \cite{Alvertis2023PRL, Alvertis2020PRB}, charge separation, internal conversion \cite{Jiang2021}, and recombination. Light exposure in perovskites can induce its degradation \cite{Hong2019}, phase transitions \cite{Leonard2023}, rotations of the organic cation \cite{Wu2017}, and change of their lattice parameters \cite{Tsai2018, Dubey2026, Zhang2023NatPhys, Li2021NatNano}. In reference \cite{Lan2019}, ultrafast experiments study the exciton coupling to coherent phonons in perovskites. In 2D materials, coherent phonons produced by light absorption were experimentally studied using ultrafast optical measurements \cite{Trovatello2020, Sayers2023, Li2021, Jeong2016, Mor2021}, and stimulated Raman spectroscopy can be used to investigate time dependent light-induced changes in materials \cite{Dasgupta2020, Batignani2020JPCL}.

From the ab initio point of view, materials electronic structures are calculated using the GW method \cite{Hybertsen1986}, excitonic effects are studied using the Bethe-Salpeter Equation (BSE) \cite{Rohlfing2000}, and vibrational properties are studied using Density Functional Pertubation Theory (DFPT). Those methods are well established, although the combination of those two features is just being performed by recent works through the investigation of exciton-phonon coupling (EPC) influence on optical absorption and emission \cite{Paleari2019, Cannuccia2019, Huang2021, Shree2018, Chen2020, Antonius2022, Zanfrognini2023PRL, Lechifflart2023, Paleari2022, Brem2020, Libbi2022, Huang2021, Alvertis2020PRB, Mishra2019, Kelley2019}. EPC redshifts the optical absorption at finite temperature \cite{Huang2021, Chan2023, Paleari2022, Mishra2019}, increases the peaks linewidth \cite{Shree2018, Chen2020, Chan2023, Paleari2022, Mishra2019, Vuong2017} and makes new satellite peaks appear with energy higher than the main peak \cite{Antonius2022}. Exciton energy is redshifted due to the Fan-Midigal like exciton self-energy \cite{Chan2023, Antonius2022}. In references \cite{Reichardt2020, Reichardt2019} exciton-exciton scattering mediated by phonons is included in \textit{ab initio} theoretical Raman spectra calculations, while in reference \cite{Gillet2017} excitonic effects are included in second-order resonant Raman scattering calculations, which are important to understand Raman spectra of 2D TMDs \cite{Wang2018}. In reference \cite{Geondzhian2018} it was shown that EPC is necessary to interpret experimental resonant inelastic x-ray scattering results. EPC is also essential to understand exciton relaxation pathways \cite{Cohen2024, Chen2020, Chan2023, Jiang2021}. In references \cite{Alvertis2023, Filip2021} the electron-hole interaction is modified by phonon screening in BSE calculations, which successfully describes the exciton binding energy dependence on temperature.

Interaction between excitons and phonons in \textit{ab initio} has generally been treated via finite differences, also known as “frozen phonon” calculations, in which separate calculations are done on a series of structures with small atomic displacements, and the differences are used to approximate a derivative \cite{Paleari2019, Cannuccia2019, Gillet2013, Gillet2017}. 
Performing the complete set of $3N$  calculations ($N$ is the number of atoms) is computationally prohibitive except for small systems \cite{aylak2021} as BSE already scales as $O(N^6)$. A recent linear-response cumulant approach to exciton-phonon coupling effects on absorption spectra \cite{Cudazzo2020} has been shown for a model system. In references \cite{Filip2021, Adamska2021} phonon effects are included in the kernel of the Bethe-Salpeter Equation, which increased the screening on the excitons and reduced their binding energy. In ref. \cite{Chen2020} the authors derive the exciton-phonon coupling by writing the BSE hamiltonian on phonon-perturbed basis and project it in a perturbed basis where the BSE solutions are known. In references \cite{Dai2024PRB, Dai2024PRL} a novel approach that combines solutions from the BSE and DFPT is presented, where the lattice distortions due to the EPC are obtained by minimizing the total energy composed as a sum of the excitation energy and the DFT energy in the harmonic approximation. This approach can be used to study excitons with polaronic character and is based on previous works that studied polarons \cite{Sio2019PRB, Sio2019PRL}. Perebeinos et. al \cite{Perebeinos2005} used a similar approach combining tight binding and classical force fields to study EPC in carbon nanotubes. In Non-Adiabatic Molecular Dynamics (NAMD) simulations, excitonic effects can be added through solutions of GW-BSE calculations on ensembles of atomic trajectories \cite{Zheng2019, Jiang2021}.   

\begin{figure}[ht]
\includegraphics[width=\columnwidth]{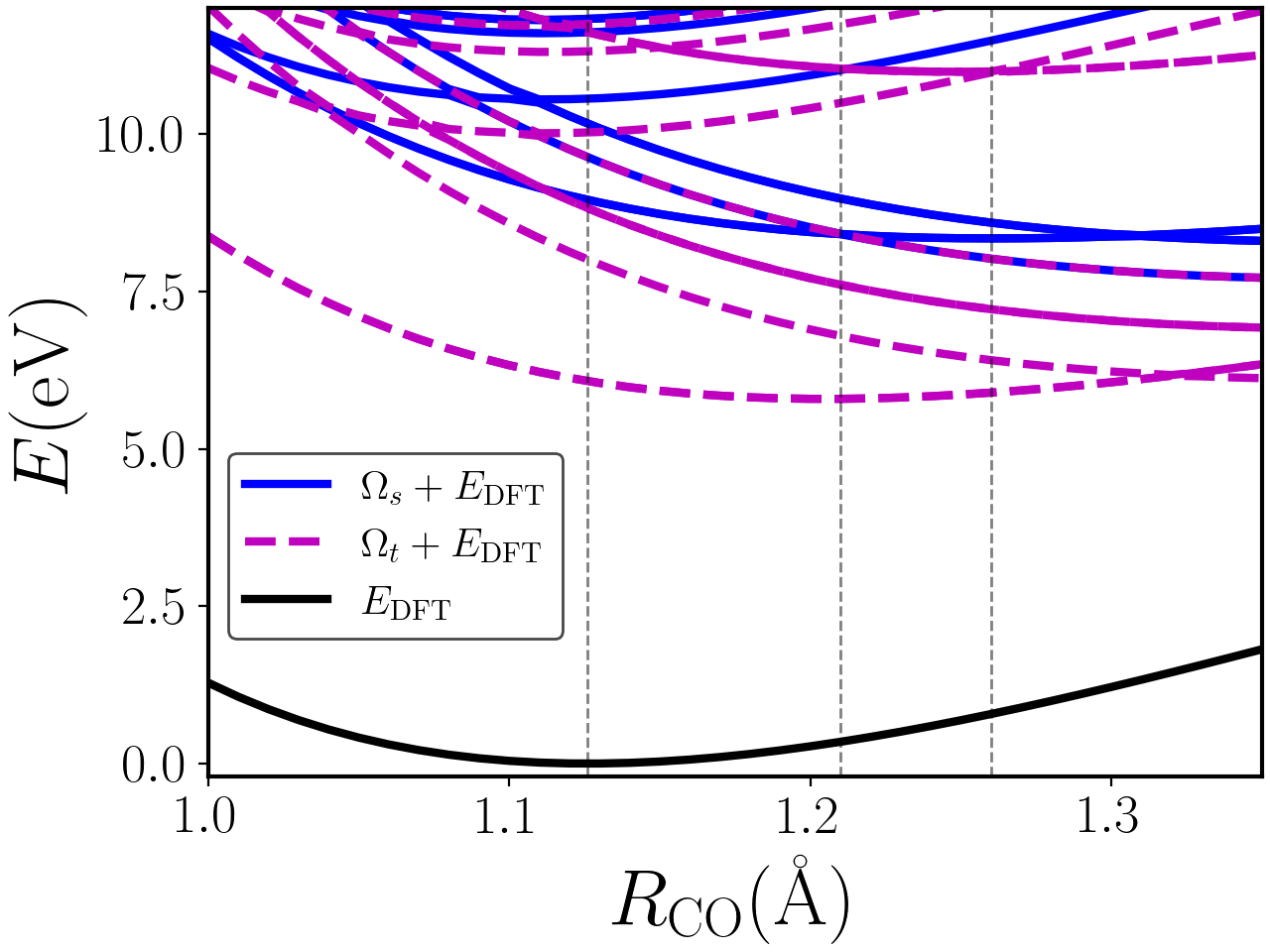}
\caption{Ground state energy calculated at DFT level (black line), excited state energies calculated for singlet (blue line) and triplets (purple line). Vertical lines correspond to equilibrium distances 1.12 \text{\AA}, 1.21 \text{\AA}, and 1.26 \text{\AA}, corresponding to ground, excited triplet, and singlet, respectively.}
\centering
\label{fig:PES_CO}
\end{figure}

Besides the efforts of understanding EPC, there are just a few works that explore the molecular mechanisms of self-trapping of excitons using \textit{ab initio} methods \cite{Wang2019, Luo2018, IsmailBeigi2005, Dai2024PRB, Dai2024PRL}.  In those works the excitation energy is calculated as a function of the collective displacements of atoms and the exciton trapping is pointed out when the total energy surface reaches a minimum. As an example, we show the case for the CO molecule in figure \ref{fig:PES_CO}, where we show the excited state energy of singlet and triplet excitons. As the CO molecule has just one degree of freedom then one can study easily the evolution of its excited state surfaces as a function of the CO bond length. For both singlet and triplet lowest energy excitons the minimum of the total energy leads to an increase of the CO bond length (see table \ref{tab:equilibrium_distance_excited_states_CO}). 

In more complex cases, one needs to find the minimum of the $3(N-1)$ dimensional excited state surface. One can use any relaxing algorithm that samples the exciton energy $\Omega(R)$ in different configurations $R$, although this would demand several GW-BSE calculations. One useful information in this relaxation process is the gradient of the exciton energy with respect to the atomic coordinates $\nabla_R \Omega (R)$. This allows the use of algorithms like Steepest Descent (SD) and Conjugate Gradient (CG), highly used in \textit{ab initio} simulation packages. In this paper, we focus on the gradient of the exciton energy obtained from GW-BSE calculations. GW-BSE is the state of art to study excitons in materials \cite{Rohlfing2000, Deslippe2012, DelBen2024} with excellent agreement with experiments. Our approach combines results from GW-BSE and Density Functional Perturbation Theory (DFPT) \cite{Giannozzi2009, Giannozzi2017, Giannozzi2020, Baroni2001, Giustino2017}. We call the negative of this gradient as Excited State Forces (ESF). In relaxation processes, we minimize the total energy, which is the sum of the ground state energy and the excitation energy $E_{\rm{T}} =E_{\rm{GS}} + x \Omega $, for a given exciton volumetric concentration $x$, until the total force $F_{\rm{T}} =F_{\rm{GS}} - x \nabla \Omega $ is zero.

\section{Overview}

The paper is organized in the following way: in section \ref{sec:theoretical_background} we introduce the theory of excited state forces. In section \ref{sec:code_implementation} we discuss details of our implementation. In section \ref{sec:ASR_on_ELPH_coeffs} we discuss and analyze the importance of acoustic sum rules over the ELPH coefficients. In section \ref{sec:benchmark_CO} we calculate ESF for the CO molecule and compare it with finite difference results. We also discuss the performance of approximations introduced in section  \ref{sec:theoretical_background}. On section \ref{sec:benchmark_LiF} we analyze the ESF for LiF and then discuss about different excited state relaxation strategies in section \ref{sec:relaxation_schemes}. We studied self-trapped excitons and polaronic excitons similar to the hole excitonic polarons of references \cite{Dai2024PRB, Dai2024PRL}. On section \ref{sec:exc_ph_symm} we project excited state forces in a phonon displacement basis for monolayer MoS$_2$ and analyze exciton-phonon symmetries and in section \ref{sec:ESF_and_SOC} we analyze the effect of spin-orbit coupling in excited state forces. Finally in section \ref{sec:conclusions} we present our conclusions. Sections \ref{sec:ESF_equations} and \ref{sec:renormalization_ELPH_coefficients_equations} we provide detailed mathematical derivation of of equations used in the text. In section \ref{sec:computational_details} we provide computational details of our calculations and in section \ref{sec:tight_binding_model_H2} we write a simple tight binding model for a $\rm{H}_2^+$ molecule, explaining why the ESF in the CO molecule is repulsive.

\section{Analytical Excited State Forces Theory}
\label{sec:theoretical_background}

{\c{C}}aylak and Baumeier studied the relaxation of the excited state of small molecules \cite{aylak2021} by performing GW/BSE calculations for different atomic configurations. Gradients of the excited state energies were obtained using finite differences, and in the case of large molecules with more degrees of freedom, just displacements that preserved the molecule symmetry were performed. 

At TDDFT level, excited state forces can be calculated analytically \cite{Tsukagoshi2012, Andrade2015, Sitt2007, Haruyama2012, Jin2023, Jin2024, Furche2002}. They do not need approximations beyond the ones used in TDDFT (i.e. linear response) and depend on the quality of the exchange-correlation functional \cite{Sitt2007}. TDDFT works well in molecules, but standard DFT functionals do not reliably predict optical excitations in solids or nanostructures, because they lack the necessary long-range behavior to describe excitonic interactions \cite{Onida2002, Wing2019}. So excited state forces calculated from GW-BSE results show the same advantages that GW-BSE calculations show in comparison to TDDFT. 

In reference \cite{Kirchhoff2024} a combination of constrained DFT (cDFT) and GW-BSE calculations is used. The excited states are relaxed at cDFT level, and once a configuration change path is obtained then exciton energies are calculated at GW-BSE level to refine the position of the minimum of the excited state energy surface. The cDFT relaxations do not include excitonic effects, just change the occupations in bands, although this approach may give good results if the studied exciton is well represented by the transition given by cDFT \cite{Kirchhoff2024}.


In 2003 Ismail-Beigi and Louie published an approach to calculate forces for the excited states in molecules and solids by combining solutions of the BSE and electron-phonon (ELPH) coefficients from Density Functional Perturbation Theory (DFPT) \cite{IsmailBeigi2003}. For the best of our knowledge, just one published work from the same authors used this approach in the last 20 years to study exciton self-trapping in SiO$_2$ \cite{IsmailBeigi2005}. As GW/BSE calculations became more common in the materials science community \cite{Deslippe2012, DelBen2024}, we call attention to this method to study the relaxation of excited states.  We revisit this theory and also analyze its convergence and performance in detail.

The basic concept in the excited state forces is that after absorbing light and creating an exciton $|A\rangle$ with energy $\Omega^A$ the total energy for a given system is given by $E=E_0 + x \Omega^A$ where $E_0$ is the ground state energy and $x$ is the exciton volumetric concentration (i.e. excitons per unit cell). For ground-state calculations, we are using Density Functional Theory using the Quantum Espresso package \cite{Giannozzi2009, Giannozzi2017, Giannozzi2020} and for the excited states we are using the BerkeleyGW code \cite{Deslippe2012, Hybertsen1986, Rohlfing2000, DelBen2024}. Computational details are provided in section \ref{sec:computational_details}. This total energy depends parametrically on the atomic positions $R$, which defines a $3N$ dimensional potential energy surface ($N$ being the total number of atoms). The excited state force is given as the negative gradient with respect to the atomic positions of the exciton energy $- \nabla \Omega^A$  and the total force is the sum of the excited state force times the exciton concentration $x$ and the ground state force, which is well established in DFT codes. In reference \cite{VillalobosCastro2023} derivatives of the BSE hamiltonian with respect to the electric field were used to calculate dipole moments in a similar approach to the one used here. More recently, two new ESF works appeared based on the GWL \cite{JinArxiv2026} and WEST \cite{Alrahamneh2025} codes, with results for CO bond length in the first singlet excited state similar to ours (see table \ref{tab:equilibrium_distance_excited_states_CO}). This agrees with the fact that excited state forces are based on previous many-body and electron-phonon calculations, and therefore reliable excited state forces can be obtained easily, once one already has converged excitonic and electron-phonon results. We already used our implementation to study exciton-phonon interactions in 1D perovskites \cite{Karkee2026AdvMatTech} and light-induced interlayer distance increase in twisted bilayer WSe$_2$ \cite{DelGrande2026arxiv}, both works in agreement with experimental data \cite{Yuan2017, NakamuraAdvMat2026}.

The BSE hamiltonian in the Tamm-Dancoff approximation is composed of two parts: an independent particle transition $D_{\mathbf{k}cv,\mathbf{k}'c'v'}=(E^{\rm{QP}}_{c\mathbf{k}}-E^{\rm{QP}}_{v\mathbf{k}})\delta_{\mathbf{k}\mathbf{k}'}\delta_{cc'}\delta_{vv'}$, where $E^{\rm{QP}}_{i\mathbf{k}}$ is the quasi-particle energy of the state $|i\mathbf{k}\rangle$, obtained in the GW approximation \cite{Hybertsen1986, DelBen2024, Deslippe2012}. $D$ is diagonal in the basis of the transitions $\mathbf{k}v\to \mathbf{k}c$ basis (represented by the ket $|\mathbf{k}cv\rangle$). The second part that composes the BSE Hamiltonian is the kernel $K^{\rm{eh}}$ that contains the electron-hole interaction \cite{Rohlfing2000, DelBen2024, Deslippe2012}. This kernel can be either a singlet or a triplet kernel. 

In section \ref{sec:ESF_equations} we provide detailed derivation of ESF expressions. The ESF expression used in reference \cite{IsmailBeigi2003} comes from writing the exciton energy as $ \Omega = \sum_{\mathbf{k}cv \mathbf{k}'c'v'} A_{\mathbf{k}cv} A^*_{\mathbf{k}'c'v'}  \langle \mathbf{k}cv|H^{\rm{BSE}}|\mathbf{k}'c'v'\rangle $, and calculating the gradient of the BSE matrix elements, ignoring derivatives of the exciton coefficients $A_{\mathbf{k}cv}$. The full expression is given by  
\begin{equation}
\begin{split}
    \vec{F}^A & = 
    - \sum_{\mathbf{k}cv\nu} |A_{\mathbf{k}cv}|^2 (g_{\mathbf{k}c,\mathbf{k}c}^{\rm{DFT},\nu} - g_{\mathbf{k}v,\mathbf{k}v}^{\rm{DFT},\nu}) \hat{\nu}  \\
    & - \sum_{\mathbf{k}cv \mathbf{k}'c'v' \nu} A_{\mathbf{k}cv} A^*_{\mathbf{k}'c'v'} \nabla_\nu \langle \mathbf{k}cv | K^{\rm{eh}} | \mathbf{k}'c'v' \rangle \hat{\nu}
\end{split}
\label{eq:ibl_esf_eq_full}
\end{equation}
where $\hat{\nu}$ is one of the displacement pattern (i.e. a phonon mode), $g^{\rm{DFT},\nu}_{\mathbf{k}i,\mathbf{k}'j}=\langle \mathbf{k}i|\nabla_\nu H^{\rm{DFT}}|\mathbf{k}j\rangle$ are the electron-phonon coefficients due to the collective atomic displacement $\nu$ and $A_{\mathbf{k}cv} = \langle \mathbf{k}cv | A \rangle$ is the exciton coefficient in the basis $|\mathbf{k}cv \rangle $. The first term in equation \ref{eq:ibl_esf_eq_full} combines directly exciton coefficients with diagonal ELPH matrix elements. To evaluate the second term one needs to apply the product rule and rewrite the ket $|\mathbf{k}cv\rangle = |\mathbf{k}c\rangle \otimes |\mathbf{k}v\rangle $. Using the product rule, the second term is expanded into five terms (see section \ref{sec:ESF_equations}): one containing the gradient of the kernel operator $\langle \mathbf{k}cv |\nabla_\nu K^{\rm{eh}} | \mathbf{k}'c'v' \rangle$ and four containing the gradient of the the single particle states (see equation \ref{eq:expansion_kernel_derivatives_ibl}). The gradient of the single-particle states are expanded with first order perturbation theory
\begin{equation}
    |\nabla_\nu \mathbf{k}n\rangle = \sum_{m \neq n} \frac{\langle \mathbf{k} m | \nabla_\nu H^{\rm{DFT}} | \mathbf{k}n\rangle }{E^{\rm{DFT}}_{\mathbf{k}n} - E^{\rm{DFT}}_{\mathbf{k}m}} |\mathbf{k} m\rangle 
\end{equation}
where one can identify the numerator of the above expression as ELPH coefficients obtained from DFPT. In this expansion, conduction (valence) states are not mixed with valence (conduction) states, which is a reasonable approximation for semiconductors and insulators and is consistent with the Tamm-Dancoff approximation. 

In reference \cite{IsmailBeigi2003}, two possibilities for ESF were used: one with just the first term of equation \ref{eq:ibl_esf_eq_full} and other including the both terms and approximating the kernel inner gradients to be zero ($\langle \mathbf{k}cv | \nabla_\nu K^{eh} | \mathbf{k}'c'v'\rangle =0$). The kernel operator is composed by two parts: the direct and the exchange kernels. The exchange kernel operator does not depend on the atomic positions, so its gradient is zero. The direct kernel is the screened coulomb interaction between the electron and hole, and involves the gradient of the dielectric function, so this approximation considers that the dielectric function changes due to atomic displacements are negligible. In \cite{IsmailBeigi2003}, calculations of ESF for the CO molecule just including derivatives of the independent particle transition part missed finite difference results by about 4 eV/\text{\AA}, while when derivatives of the kernel were included the forces in the C and O atom were closer to finite differences, but differ from each other by about 2 eV/\text{\AA}, with a net non-zero force on the CO molecule center of mass. We revisited this method, implemented it in a workflow that combines results from DFPT and GW/BSE and addressed the problems faced by Ismail-Beigi and Louie. Our two major accomplishments in our methodology are: 1. detect the source of the problem of non-zero net forces on the system center of mass (see section \ref{sec:ASR_on_ELPH_coeffs}) and 2. renormalizing ELPH obtained from DFPT using our scheme and including off-diagonal ELPH coefficients (see figure \ref{fig:derivative_Omega_several_methods}).  

In contrast to Ismail-Beigi and Louie's work, we use the Hellman-Feynman theorem to calculate ESF, $\nabla \Omega^A = \langle A | \nabla H^{\rm{BSE}} | A \rangle$, as $\Omega^A$ and $|A\rangle$ are eigenvalue and eigenvector of the BSE hamiltonian $H^{\rm{BSE}}$, respectively. As shown, in section \ref{sec:benchmark_CO}, our method (eq. \ref{eq:excited_state_forces_our_method}) and Ismail-Beigi and Louie's method (eq. \ref{eq:ibl_esf_eq_full}), after our corrections, provide essentially the same results. Our implementation is simpler and avoids extra summations over conduction and valence states, which accelerates calculations of ESF. One can identify $\langle A | \nabla_\nu H^{\rm{BSE}} | A \rangle $ as the exciton-phonon coupling for excitons coupled to a zero-momentum phonon (or a linear combination of phonons with $q=0$) \cite{Antonius2022, Chen2020, Cudazzo2020, Reichardt2020, Wang2018, Zanfrognini2023PRL, Lechifflart2023, Cohen2024, Paleari2022}. 

We use the following expression for the ESF

\begin{equation}
\begin{split}
    \vec{F}^A = & - 
    \sum_{\nu \mathbf{k} cv c'v'} A_{\mathbf{k}cv} A^*_{\mathbf{k}c'v'} 
    (g_{\mathbf{k}c,\mathbf{k}c'}^{\rm{QP},\nu} \delta_{vv'} - g_{\mathbf{k}v,\mathbf{k}v'}^{\rm{QP},\nu} \delta_{cc'}) \hat{\nu} \\
    & -\sum_{\nu \mathbf{k} cv \mathbf{k}' c'v'} A_{\mathbf{k}cv} A^*_{\mathbf{k}c'v'} \langle \mathbf{k}cv | \nabla_\nu K^{\rm{eh}} | \mathbf{k}'c'v' \rangle \hat{\nu}.
\end{split}
\label{eq:excited_state_forces_our_method}
\end{equation}

Differently from reference \cite{IsmailBeigi2003} we consider the mixing of different conduction (valence) states by including off-diagonal electron-phonon coefficients, as the diagonal part of BSE hamiltonian becomes non-diagonal due to the atomic displacements $D_{\mathbf{k}cv,\mathbf{k}'c'v'} = (\langle \mathbf{k}c | H^{\rm{QP}} | \mathbf{k}c' \rangle - \langle \mathbf{k}v | H^{\rm{QP}} | \mathbf{k}v' \rangle) \delta_{\mathbf{k}\mathbf{k}'}$. Even though the off-diagonal QP matrix elements are zero ($\langle \mathbf{k}n | H^{\rm{QP}} | \mathbf{k}m \rangle = 0$), atomic displacements, induce the mixing between different states, which is quantified by the off-diagonal matrix ELPH matrix elements $\langle \mathbf{k}c | \partial H^{\rm{QP}} | \mathbf{k}c' \rangle$. Those off-diagonal ELPH matrix elements come naturally from the calculation of ESF using the Hellman-Feynman theorem (see section \ref{sec:ESF_equations}). Here we use the same approximation of considering the derivative of the kernel operator to be negligible, so the second term in the above equation is zero. Notice that our final expression does not need to include extra summations and the knowledge of kernel matrix elements as in equation \ref{eq:expansion_kernel_derivatives_ibl}. 

The second difference between our approach and the one used by Ismail-Beigi and Louie is the treatment of ELPH coefficients, as they use directly ELPH coefficients from DFPT calculated at DFT level. ELPH coupling obtained at DFT level is underestimated in relation to ELPH calculated at GW level \cite{Antonius2014, Li2019, Faber2011, Faber2015} (see equation \ref{eq:relation_elph_qp_dft}). Frozen phonon calculations at GW level show that ELPH energies are normalized more than 40\% for diamond \cite{Antonius2014}. Faber et. al \cite{Faber2015, Faber2011} calculated the effective electron-phonon coupling potential $V^{ep}$ and observed that EPC effective potentials using self-consistent GW agreed better with experimental data than one shot G$_0$W$_0$ and DFT results were substantially smaller than any GW calculation \cite{Faber2011}. The same trends were observed for diamond and graphene in reference \cite{Faber2015}. Li et al developed a self-consistent methodology to calculate el-ph coefficients at GW level, called GWPT, analogous to DFPT. This method had a better agreement to experimental data for the critical temperature for the superconductive phase of $\mathrm{Ba_{1-x} K_x Bi O_3}$ than DFPT, with enhancements of ELPH coefficients up to 60\% \cite{Li2019}. Splits of energy levels in diamond indicate an increase of about 15\% in ELPH coefficients \cite{Li2019}. 
GWPT at BerkeleyGW is not yet available to the public, and its computational cost is substantial larger than DFPT. We use an alternative approach to compute ELPH matrix elements at GW level at low computational cost and larger than ELPH matrix elements obtained at DFT level. 

The electron-phonon coefficients at QP level relate to the ones calculated at DFT level by
\begin{equation}
    \langle \mathbf{k}i | \partial H^{\rm{QP}} | \mathbf{k}j \rangle = \langle \mathbf{k}i | \partial H^{\rm{DFT}} | \mathbf{k}j \rangle + \langle \mathbf{k}i | \partial (\Sigma - V_{xc}) | \mathbf{k} j \rangle
    \label{eq:relation_elph_qp_dft}
\end{equation}
, where $\Sigma$ is the self-energy operator and $V_{xc}$ is the exchange-correlation functional \cite{Hybertsen1986}. This task can be done either using FD or GWPT \cite{Li2019}. In several works \cite{IsmailBeigi2003, VillalobosCastro2023, Chen2020, Reichardt2020}, electron-phonon coefficients are calculated directly from DFPT calculations, neglecting the derivatives of $( \Sigma - V_{xc} )$. We chose to use a renormalization scheme that is based on the approximation that the eigenvectors from the DFT Hamiltonian are also eigenvectors of the QP Hamiltonian, which is the essence of the successful one-shot GW approximation \cite{Hybertsen1986} (see section \ref{sec:renormalization_ELPH_coefficients_equations}). This scheme is given by \cite{Strubbe2012, Levine1989}

\begin{widetext}
\begin{equation}
    \langle i | \partial H^{\rm{QP}} | j \rangle  = 
    \begin{cases}
\left( \frac{E^{\rm{QP}}_i - E^{\rm{QP}}_j}{E^{\rm{DFT}}_i - E^{\rm{DFT}}_j} \right)  \langle i | \partial H^{\rm{DFT}} | j \rangle , & \text{if } E^{\rm{DFT}}_i \neq E^{\rm{DFT}}_j \\
\langle i | \partial H^{\rm{DFT}} | j \rangle , & \text{else.} 
    \end{cases}
    \label{eq:renormalization_elph_coeffs}
\end{equation}
\end{widetext}

The diagonal matrix elements remain the same while off-diagonal matrix elements are renormalized based on QP and DFT energy level differences. We explore the performance of this approximation in sections \ref{subsec:ELPH_derivatives_diagonal} and \ref{subsec:ELPH_derivatives_offdiagonal}.


One practical problem that Ismail-Beigi and Louie faced, is that in calculations for the CO molecule including derivatives of the kernel, the ESF on the C and O atom did not obey Newton's third law, and the net force on the CO molecule was about 2 eV/\text{\AA} instead of zero \cite{IsmailBeigi2003}. One solution to this issue was to subtract the force on the center of mass after the ESF calculation. They attributed this fact to the approximation $\delta W / \delta G \approx 0$ ($W$ is the screened coulomb interaction and $G$ the Green's function used in the GW approximation). We noticed that actually this issue is caused intrinsically by the not obedience of the acoustic sum rule by the ELPH coefficients from DFPT calculations in Quantum Espresso code. Therefore the approximation  $\delta W / \delta G \approx 0$ can still be safely used. We solved this problem by applying an Acoustic Sum Rule (ASR) to the ELPH coefficients before the ESF calculations (see section \ref{sec:ASR_on_ELPH_coeffs}). Both our and Ismail-Beigi and Louie's method, after applying the ASR on the ELPH coefficient, presented results where the ESF on the center of mass is zero. 

\section{Code implementation}

\label{sec:code_implementation}

\begin{figure*}
    \centering
    \includegraphics[width=1\linewidth]{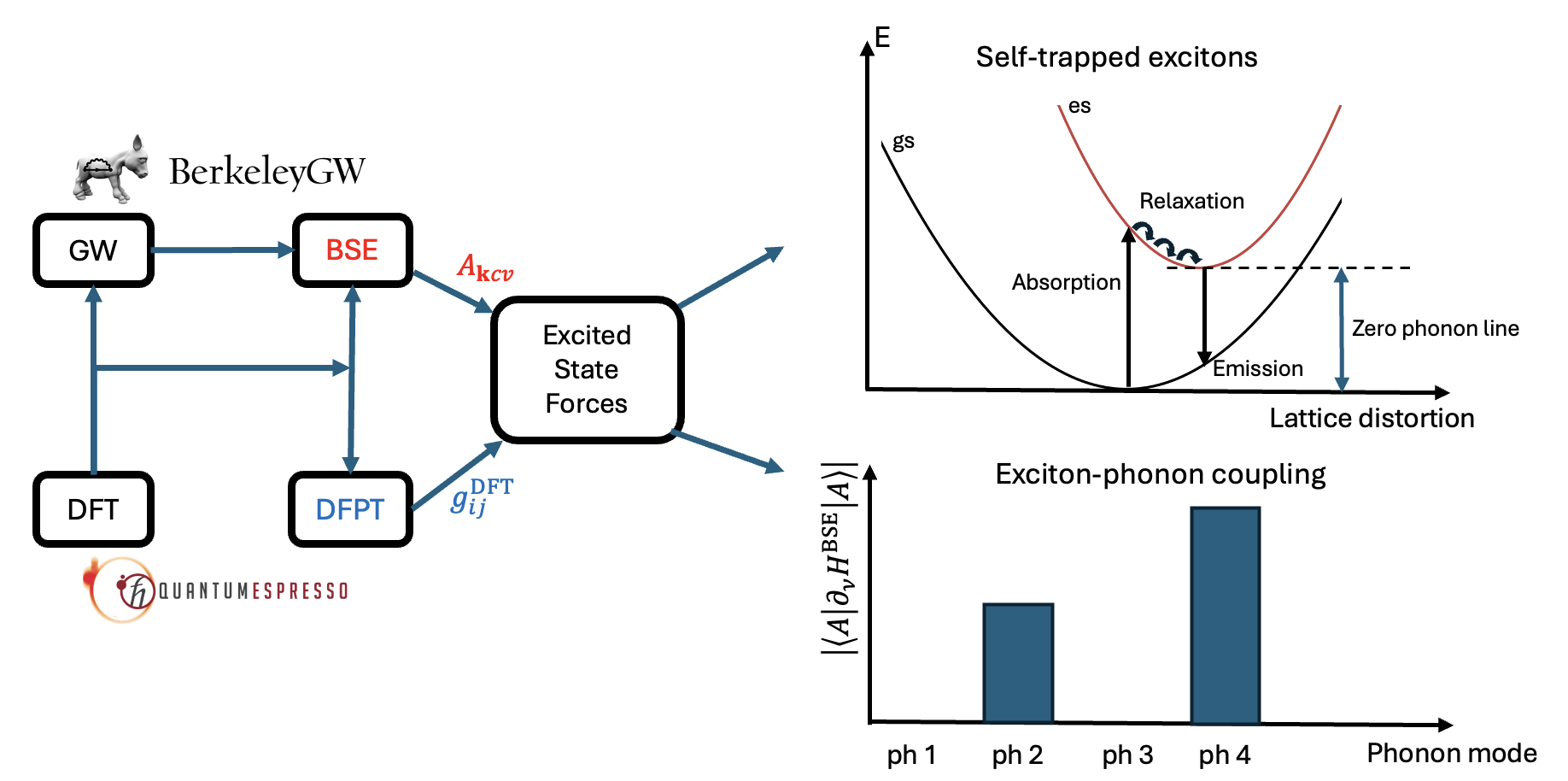}
    \caption{Diagram of the ESF workflow that combines electron phonon matrix elements from DFPT and exciton coefficients from BSE}
    \label{fig:diagram_workflow_ESF}
\end{figure*}

Our current implementation was done using Python and is available in \cite{github_esf} with several application examples. Our code reads ELPH coefficients in \texttt{XML} format obtained from DFPT from Quantum Espresso (QE) code \cite{Giannozzi2009, Giannozzi2017, Giannozzi2020} and exciton coefficients in \texttt{HDF5} format from BerkeleyGW code \cite{Deslippe2012}, perform numerical operations using \texttt{NumPy} and can be run in parallel using the module \texttt{multiprocessing} when computing multiple exciton-phonon matrix elements. A diagram illustrating the workflow to obtain ESF is shown on Fig. \ref{fig:diagram_workflow_ESF}. The wavefunctions that are used to calculate the ELPH coefficients must be the same used to create the BSE Hamiltonian, so the coefficients $A_{\mathbf{k}cv}$ and $g^{\mu}_{\mathbf{k}i,\mathbf{k}j}$ are consistent regarding their complex phases \cite{Alvertis2023}. In the BerkeleyGW code \cite{Deslippe2012} general workflow, one computes the quasi-particle energies (GW approximation) and kernel matrix elements $\langle \mathbf{k}cv | K | \mathbf{k}'c'v' \rangle$ in a coarse $k$-grid and then interpolates those quantities to a fine $k$-grid where the BSE hamiltonian is built. We work with two possibilities: 1. the DFPT coefficients are calculated in the fine grid directly and 2. the DPFT are calculated in the coarse grid and one can use interpolation coefficients calculated by the BerkeleyGW code to compute the DFPT coefficients in the fine grid, but other schemes can also be used to interpolate ELPH coefficients \cite{Noffsinger2010, Ponc2016, Lee2023, ZhouCPC2021}. Current implementations of QE (version 7.3) just calculates ELPH of semiconductors and insulators by including smearing \cite{Jorgensen2021ModSimMat}

In general, ESF calculations with our code take about a few minutes in a personal computer. We also implemented ESF for exciton with finite-momentum $Q$ \cite{QiuPRL2015}. Our code is also able to calculate off-diagonal exciton-phonon matrix elements $\langle A |\nabla_\nu H^{\rm{BSE}}|B\rangle $ by replacing $A_{\mathbf{k}cv}$ in equation \ref{eq:excited_state_forces_our_method} by $B_{\mathbf{k}cv}$ the exciton coefficients of exciton $B \neq A$. Those off-diagonal matrix elements can be used with Fermi golden rule to calculate transition rates from exciton $A$ to $B$ induced by the phonon mode $\nu$ and are used to calculate Raman spectra with excitonic effects \cite{Reichardt2020, Reichardt2019} and exciton relaxation times \cite{Chen2020}. By now, our implementation just use phonons with no momentum, so one can just calculate exciton transitions between excitons with the same momentum. 

\section{Acoustic Sum Rule on ELPH coefficients}

\label{sec:ASR_on_ELPH_coeffs}

\begin{figure}[ht]
\includegraphics[width=\linewidth]{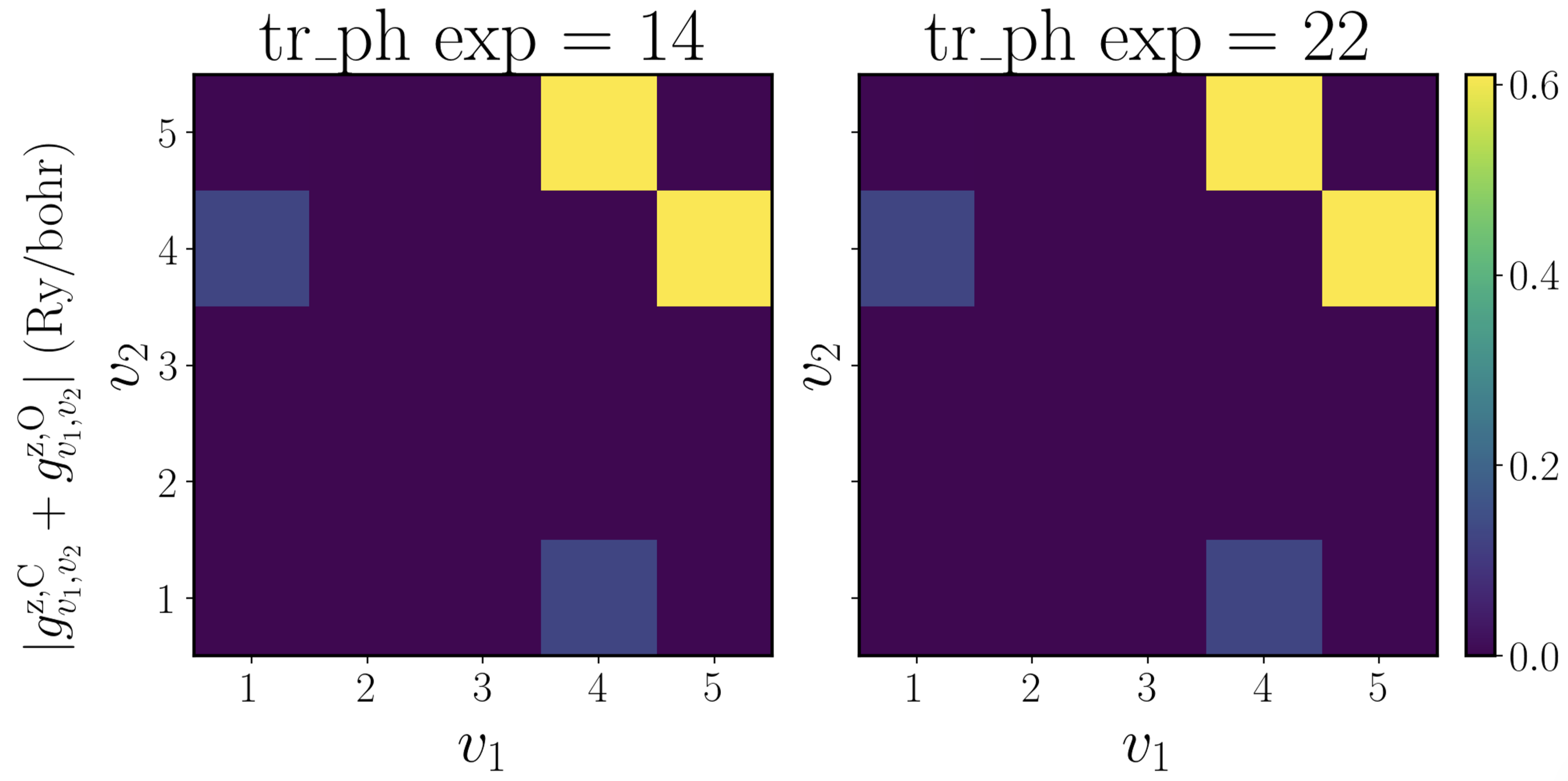}
\caption{Sum of electron-phonon coefficients due to displacements parallel to CO molecule bond of oxygen and carbon atoms using a threshold of $10^{-14}$ (left) and $10^{-22}$ (right) for the self-consistencey of derivatives in DFPT. Both summed represent an acoustic mode for which this sum should be zero. For some off-diagonal ELPH matrix elements this sum is not zero, regardless of the convergence of parameters used in DFPT calculations, and therefore our ASR (eq. \ref{eq:ASR})}
\centering
\label{fig:sum_g_z_dir_RCO}
\end{figure}

In reference \cite{IsmailBeigi2003} the excited state forces using the kernel on the C and O atoms are different from each other. They attributed this to their approximation $\partial W \approx 0$. We also faced this problem when using equation \ref{eq:excited_state_forces_our_method}. In our investigations, we calculated the sum of electron-phonon coupling coefficients given by

\begin{equation}
    S^n_{ij} = \sum_\nu \langle i | \partial_\mu H^{\rm{DFT}} | j \rangle ( \hat{n} \cdot \hat{\nu}) 
    \label{eq:ASR}
\end{equation}

\noindent where $\hat{\nu}$ is the $3N$ component unitary vector parallel to the phonon mode displacement and $\hat{n}$ is the $3N_{\rm{atoms}}$ component vector that represents the system center of mass displacement in the cartesian $x$, $y$ or $z$ directions. This quantity should be zero, as a pure center of mass displacement should have $\partial V_{scf} = 0$. We noticed that it was not obeyed in the case of the CO molecule, even when performing those test calculations with different pseudopotentials and precision thresholds. We also checked that this was not a convergence issue as shown in figure \ref{fig:sum_g_z_dir_RCO}. Our solution to this was to apply an Acoustic Sum Rule (ASR) to impose the sum in equation \ref{eq:ASR} to be zero, by making $g^n_{ij} \to g^n_{ij} -S^n_{ij}/N_{\rm{atoms}}$. We verified that this has the same effect as imposing that the force on the center of mass is zero, as it is done in reference \cite{IsmailBeigi2003}.

\section{Excited state forces vs Finite Differences in CO}
\label{sec:benchmark_CO}

\begin{figure}[ht]
\includegraphics[width=\columnwidth]{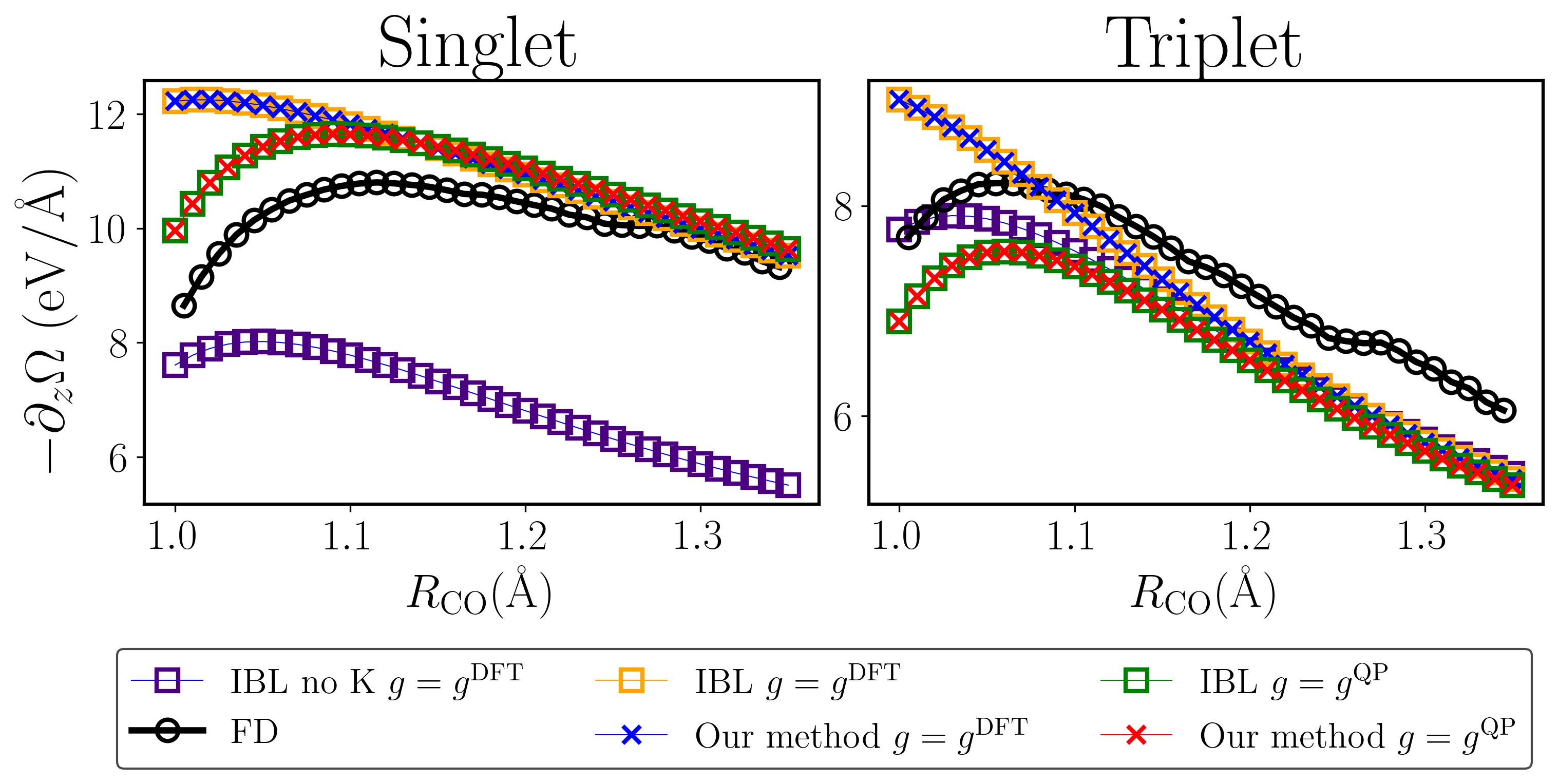}
\caption{Excited state forces for the singlet (left) and triplet (right) excitons in CO as function of $R_{\rm{CO}}$ bond length using different methods. Our reference data is Finite Differences (FD) in black circles. Singlet results using Ismail-Beigi and Louie's (IBL) method (eq. \ref{eq:ibl_esf_eq_full}) not including kernel derivatives deviate from FD by $\sim$ 4 eV/\text{\AA} as in reference \cite{IsmailBeigi2003}. Our method (eq. \ref{eq:excited_state_forces_our_method}) has a great agreement with IBL method when both use the same kind of ELPH coefficients. Our scheme to renormalize ELPH coefficients (eq. \ref{eq:renormalization_elph_coeffs}) improves the agreement of our ESF with FD in the region $R_{\rm{CO}} < 1.1 \text{\AA}$.}
\centering
\label{fig:derivative_Omega_several_methods}
\end{figure}

We start by analyzing excited state forces for the singlet (\(\text{A}^1 \Pi\)) and triplet (\(\text{A}^3 \Pi\)) excitons in the CO molecule and our results are summarized in Figure \ref{fig:derivative_Omega_several_methods}. The CO molecule has only one degree of freedom, so we can easily compare results that used analytical ESF with two-point finite differences (FD), which we use as our reference data. First, we note that the FD curves are not as smooth as the curves using our method and the methods in reference \cite{IsmailBeigi2003}. This is due to the limited size of the BSE hamiltonian and as we change the bond length the nature of the conduction states included in BSE Hamiltonian changes. Similarly to reference \cite{IsmailBeigi2003} ESFs using equation \ref{eq:ibl_esf_eq_full} with only the first term, disagree from FD by about 4 eV/\text{\AA}. After the use of the ASR both Ismail-Beigi and Louie's and our methods have a much better agreement with FD, and most essentially they agree to each other in a degree of their numerical precision, as it is expected, as both methods calculate  the same quantities, using the same approximations ($\langle \mathbf{k}cv | \partial K^{\rm{eh}}|\mathbf{k}'c'v'\rangle \approx 0$). We also explored the effect of using ELPH coefficients at DFT level and our renormalization scheme (eq. \ref{eq:renormalization_elph_coeffs}). The renormalization of ELPH coefficients make results agree better with FD in the region $R_{\rm{CO}} < 1.1$ \text{\AA}, while in the region $R_{\rm{CO}} > 1.1$ \text{\AA} it does not influence substantially for both singlet and triplet calculations. Therefore we can state that two sources of errors in the ESF theory used here are the approximation of the kernel derivative and a precise evaluation of the ELPH coefficients at GW level (i.e. GWPT). In table \ref{tab:equilibrium_distance_excited_states_CO} we compare the equilibrium bond length for the excited singlet and triplet states of the CO molecule with experimental data and other ab initio studies. Our results deviate from experimental values by just a few pm.

\begin{table*}[]
    \centering
    \begin{tabular}{|c|c|c|c|}
    \hline
        $R_{\rm{CO}} $(pm) & Ground State & Singlet & Triplet  \\
    \hline
      This work   & 112 & 126 & 121 \\
      Experimental \cite{NIST_page} &   & 123.53   &  120.574 \\
      TDDFT (LDA) \cite{Liu2011} &  & 122 & 120 \\
      TDDFT (PBE) \cite{Liu2011} &  & 124 & 121 \\
      TDDFT (B3LYP) \cite{Liu2011} &  & 123 & 120 \\
      TDDFT \cite{Haruyama2012} &  & 122  & \\
      CIS \cite{Liu2011}  &   &  118 & 121 \\
      GW-BSE \cite{IsmailBeigi2003} &  & 126 & \\
      GW-BSE \cite{Alrahamneh2025}  & 114 & 128    &  \\
      GW-BSE \cite{JinArxiv2026}  &  & 126 &  \\
      G$_0$W$_0$/BSE (full BSE / FF) \cite{aylak2021} &  & 124.1  & \\
      G$_0$W$_0$/BSE (full BSE/ PPM) \cite{aylak2021} &  & 124.9  & \\
      G$_0$W$_0$/BSE (TDA / FF) \cite{aylak2021} &  & 124.9  & \\
      G$_0$W$_0$/BSE (TDA / PPM) \cite{aylak2021} &  & 125.6  & \\
    \hline
    \end{tabular}
    \caption{Equilibrium bond length for triplet and singlet excited states of CO molecule. CIS stands to single-excitation configuration interaction, TDDFT calculations from reference \cite{Liu2011} were done with LDA, PBE and B3LYP functionals. In the results from reference \cite{aylak2021}, TDA refers to Tamm-Dancoff approximation, while full BSE refers to not use this approximation (include transitions $c\to c'$ and $v \to v'$). PPM stands for plasmon-pole model and FF stands for full frequency, which are methods to calculate the frequency dependent dielectric function.}
    \label{tab:equilibrium_distance_excited_states_CO}
\end{table*}

We notice that the ESF for the CO is repulsive for both singlet and triplet lowest energy excitons. Those excitons are composed by conduction states with antibonding character and valence state with bonding character. As the CO bond length increases the split between those the two energy levels of the states that compose this exciton decreases, then the ESF will be repulsive. In section \ref{sec:tight_binding_model_H2} a simple tight binding model for H$^+_2$ molecule reproduces this repulsive excited state force behavior. In one of our works \cite{DelGrande2026arxiv} we found that most excitons in bilayer WSe$_2$ lead to repulsive forces between monolayers, in agreement with experimental observations in reference \cite{NakamuraAdvMat2026}, where the interlayer distance in twisted bilayer WSe$_2$ increased by 0.2 \text{\AA} after light absorption. While in reference \cite{Chen2025AdvSci} light induced lattice expansion in 2D Dion-Jacobson perovskites, although there are works where there is a light-induced contraction of the lattice parameter \cite{Li2021NatNano}.

\subsection{Diagonal matrix elements ELPH coefficients}
\label{subsec:ELPH_derivatives_diagonal}

Now we go deeper in the approximations used in our approach.

We first analyze the diagonal matrix elements. Using the Hellman-Feynman theorem, the derivatives of the DFT energy levels are equal to the diagonal ELPH matrix elements $\partial E^{dft}_{\mathbf{k}n} = \langle \mathbf{k} n | \partial H^{dft} | \mathbf{k} n \rangle$. A common approximation used in GW calculations is that the QP and DFT hamiltonians share the same eigenvectors (the $\Sigma - V_{xc}$ operator is diagonal in the basis of the DFT eigenvectors), so in this case one can also write $\partial E^{qp}_n = \langle n | \partial H^{\rm{QP}} | n \rangle$, then the difference between both derivatives is $\partial E^{qp}_n - \partial E^{dft}_n = \langle n | \partial(\Sigma - V_{xc}) | n \rangle$. We neglect this difference and approximate $\partial E^{qp}_n \approx \partial E^{dft}_n $. This is the same as it was done in reference \cite{IsmailBeigi2003}. Further work is necessary to deal with the term $\langle n | \partial(\Sigma - V_{xc}) | n \rangle$. 

\begin{figure}[ht]
\includegraphics[width=\columnwidth]{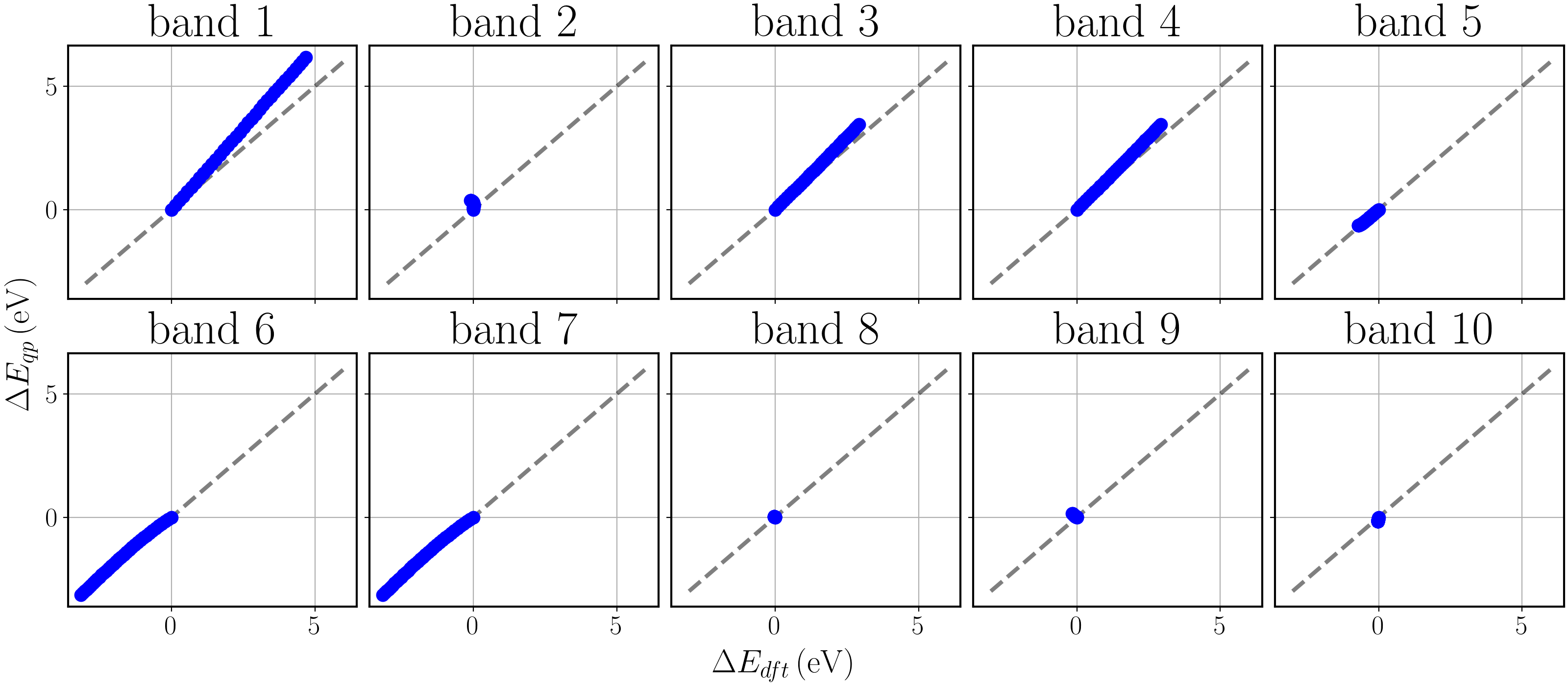}
\caption{QP energy variations vs DFT energy variations. Those energies were calculated from changing the $R_{CO}$ bond length from 1.1 to 1.35 \text{\AA} to bands 1 to 10 in CO molecule. The first five bands are full occupied. Vertical $y=0$, horizontal $x=0$ and diagonal $y=x$ dashed black lines are guides to the eye.}
\centering
\label{fig:variation_Eqp_Edft_energies}
\end{figure}

To analyze the quality of this approximation we show in figure \ref{fig:variation_Eqp_Edft_energies} the variation of the QP energy levels vs the variation of DFT energy levels when we change the CO bondlength from 1.1 to 1.35 \text{\AA}. We can see here that those two variations are correlated to each other and are, approximately, over the $y=x$ curve.

\subsection{Off-diagonal ELPH matrix elements}
\label{subsec:ELPH_derivatives_offdiagonal}

\begin{figure}[ht]
\includegraphics[width=\columnwidth]{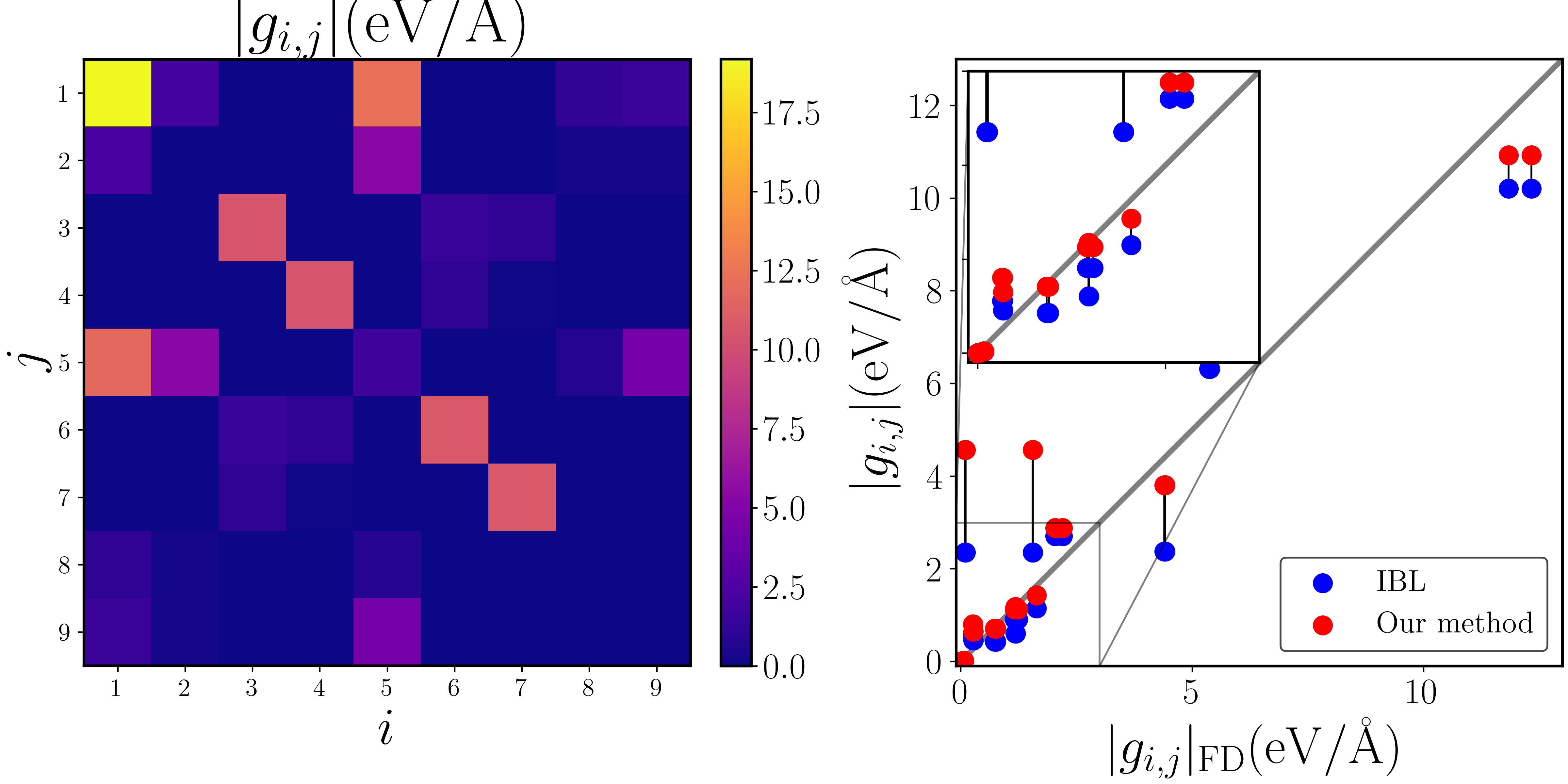}
\caption{Left: elph coefficients calculated at FD differences (see text). Right: Off-diagonal elph coefficients calculated using our renormalization scheme (red dots) and from DFPT data (blue dots) that is the approximation used in \cite{IsmailBeigi2003} as function of the elph coefficients calculated using FD. The closer the data is from the black dashed line $y=x$ the better is the agreement of the method with FD calculations.}
\centering
\label{fig:g_ij_offdiag}
\end{figure}

Now we look at the off-diagonal matrix elements derivatives. In this case, we approximate the term $\langle n | \partial(\Sigma - V_{xc}) | m \rangle$ to be $(E^{qp}_n - E^{qp}_m)/(E^{dft}_n - E^{dft}_m) - 1$, which leads to the first line of the equation \ref{eq:renormalization_elph_coeffs}. We compare this renormalization with scheme on DFPT ELPH coefficients with raw DFPT coefficients to reproduce FD calculations. The finite differences calculations were performed using the following equation

\begin{equation}
\begin{split}
    & \langle i | \partial H^{\rm{QP}} | j \rangle_{\rm{FD}} (R)  =   {\langle i | \partial H^{dft} | j \rangle}_{\mathrm{DFPT}} (R) \\ 
    &+ \frac{\langle i_R | (\Sigma - V_{xc})_{R+\delta R} - (\Sigma - V_{xc})_{R-\delta R} | j_R \rangle}{2 \delta R}
\end{split}
\label{eq:derivative_sigma_fd}
\end{equation}

\noindent where \({\langle i | \partial H^{\mathrm{dft}} | j \rangle}_{\mathrm{DFPT}} (R)\) is the ELPH coefficient calculated at DFT level using DFPT, and the second term is calculated using FD. The \((\Sigma - V_{\mathrm{xc}})\) operator is constructed with wavefunctions of CO with bond lengths \(R \pm \delta R\) ($\delta R = 10^{-4} $, \text{\AA}), and this operator is evaluated in the basis \(\vert i_R \rangle\), the state $i$ calculated for a CO bond length equal to $R$. This is done using the machinery of the BerkeleyGW code \cite{Deslippe2012}, where the $(\Sigma-V_{xc})_{R\pm\delta R}$ operator is built with the \verb|WFN_inner| and \verb|VXC| files calculated at $R\pm \delta R$ and is calculated in the basis of the \verb|WFN_outer| file, composed by states at $R$. Our results are summarized in Figure \ref{fig:g_ij_offdiag}. In fact, we observe a mean increase for ELPH coefficients by about \(35\%\) using our method in relation to ELPH from DFPT. By looking at Figure \ref{fig:derivative_Omega_several_methods}, we see that our excited state forces perform better when we use our renormalization scheme.

\subsection{Kernel derivatives}

\begin{figure}[ht]
\includegraphics[width=\columnwidth]{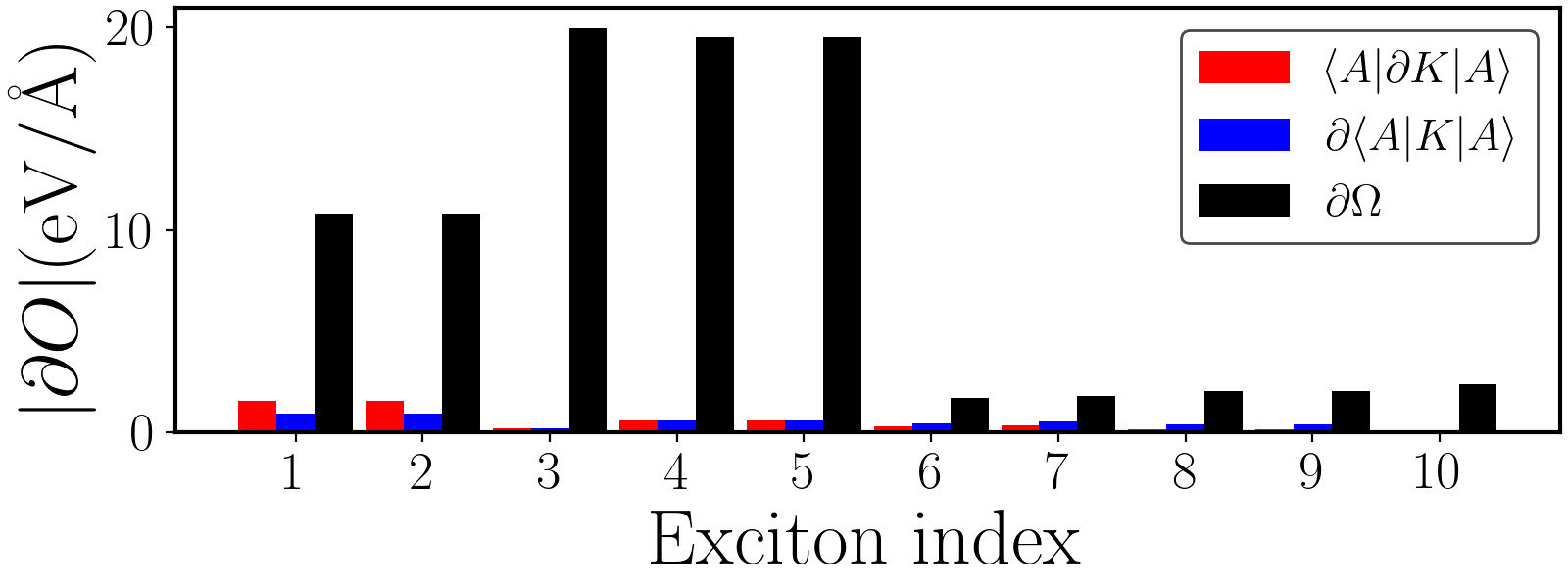}
\caption{Derivative of kernel for the ten first excitons in CO with bond length equal to 1.12\text{\AA}. In red just the operator changes, while in blue the states are allowed to change as well. In black we show the derivatives of the excitation energy. We can see that the derivatives of the kernel represent just a small fraction of the excitation energy derivative.}
\centering
\label{fig:derivative_kernel_vs_omega}
\end{figure}

In this section we explore the approximation $\langle \mathbf{k}cv | \partial K^{\rm{eh}}|\mathbf{k}'c'v' \rangle \approx 0$. We compute the derivatives $\partial \langle A|K^{\rm{eh}}|A\rangle$ and $\langle A| \partial K^{\rm{eh}}|A\rangle$ for the CO molecule with bond length equal to $R_0=1.12 \text{\AA}$. Both derivatives are evaluated with FD, where the first is given by 
\begin{equation}
\begin{split}
    \partial \langle A|K^{\rm{eh}}|A\rangle 
     = \frac{1}{2\delta R} & (
    \langle A_{R_0+\delta R}|K^{\rm{eh}}_{R_0+\delta R}|A_{R_0+\delta R}\rangle \\
    & -\langle A_{R_0-\delta R}|K^{\rm{eh}}_{R_0-\delta R}|A_{R_0-\delta R}\rangle)
\end{split}
\end{equation}
where the subscripts $\pm\delta R$ mean that the exciton coefficients $A$ and the kernel matrix were calculated at bond length $R_0 \pm \delta R$, where $\delta R= 10^{-4} \text{\AA}$. The second expression is given by
\begin{equation}
\begin{split}
    \langle A| \partial K^{\rm{eh}}|A\rangle 
     = \frac{\langle A_{R_0}|K^{\rm{eh}}_{R_0+\delta R} - K^{\rm{eh}}_{R_0-\delta R}|A_{R_0}\rangle}{2\delta R}  
\end{split}
\end{equation}
where the exciton coefficients are calculated at $R_0$. The two above expressions are not equal to each other as $|A\rangle$ is not an eigenvector of the kernel and the difference between them are the terms involving derivatives of the kets $|\mathbf{k}cv\rangle$ (see eq. \ref{eq:expansion_kernel_derivatives_ibl}). In figure \ref{fig:derivative_kernel_vs_omega} we compare those kernel derivatives with the derivatives of the excitation energy for the first ten excitons of the CO molecule, and in general the kernel derivatives are much smaller then the exciton enery derivatives, which confirms our approximation.

\subsection{Convergence of excited state forces calculations}

\begin{figure}[ht]
\includegraphics[width=\columnwidth]{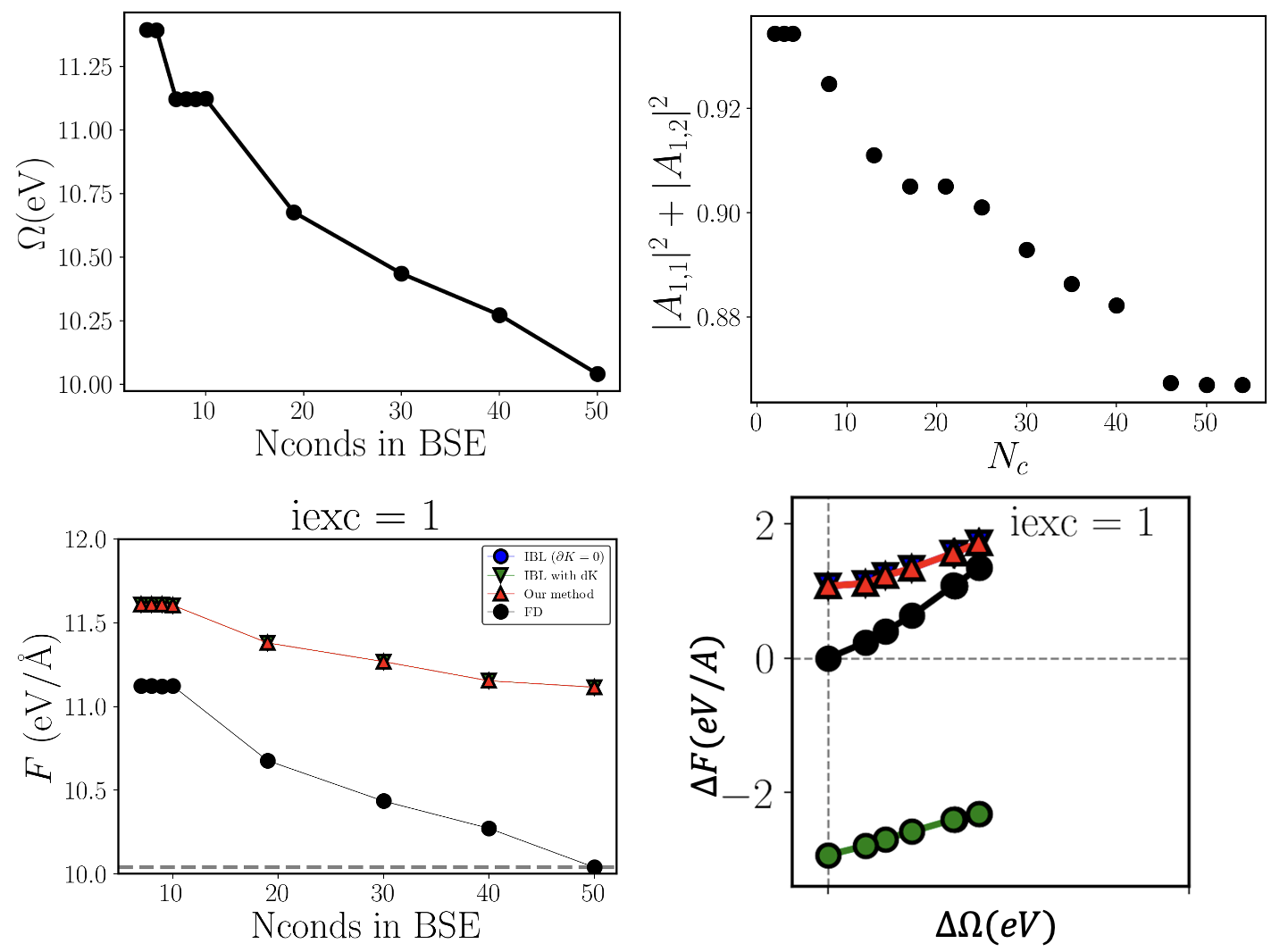}
\caption{Convergence of excitation energy, excited state forces and most important coefficients for the first exciton in CO. }
\centering
\label{fig:pannel_convergence_excited_state_forces}
\end{figure}

In BSE calculations one chooses how many conduction and valence bands will be included in the BSE Hamiltonian. This choice is made to include sufficient bands to converge the exciton energies of interest and absorption spectra. In the specific case of molecules the necessary number of conduction bands to be included can be of the order of a few hundreds \cite{Qiu2021}. Here we observe that the convergence of excited state forces follows the same tendency as the convergence of the exciton energy. As an example, we show in Figure \ref{fig:pannel_convergence_excited_state_forces} the convergence of the energy, excited state forces, and most important $A$ coefficients for the lowest singlet exciton in CO with bond length equal to 1.12 \text{\AA}. For this exciton the most important transitions are $v_1 \to c_1$ and $v_1 \to c_2$ ($c_1$ and $c_2$ here are degenerate). As the excited state forces and excitation energies depend on the $A_{\mathbf{k}cv}$ coefficients, then we observe that those three quantities follow the same trend with respect to the number of conduction bands in the BSE. In Figure \ref{fig:pannel_convergence_excited_state_forces}.d we show that the variation of the excitation energy and the variation of the excited state force from our most converged results follow a linear relation with each other.

\begin{figure}
    \centering
    \includegraphics[width=\linewidth]{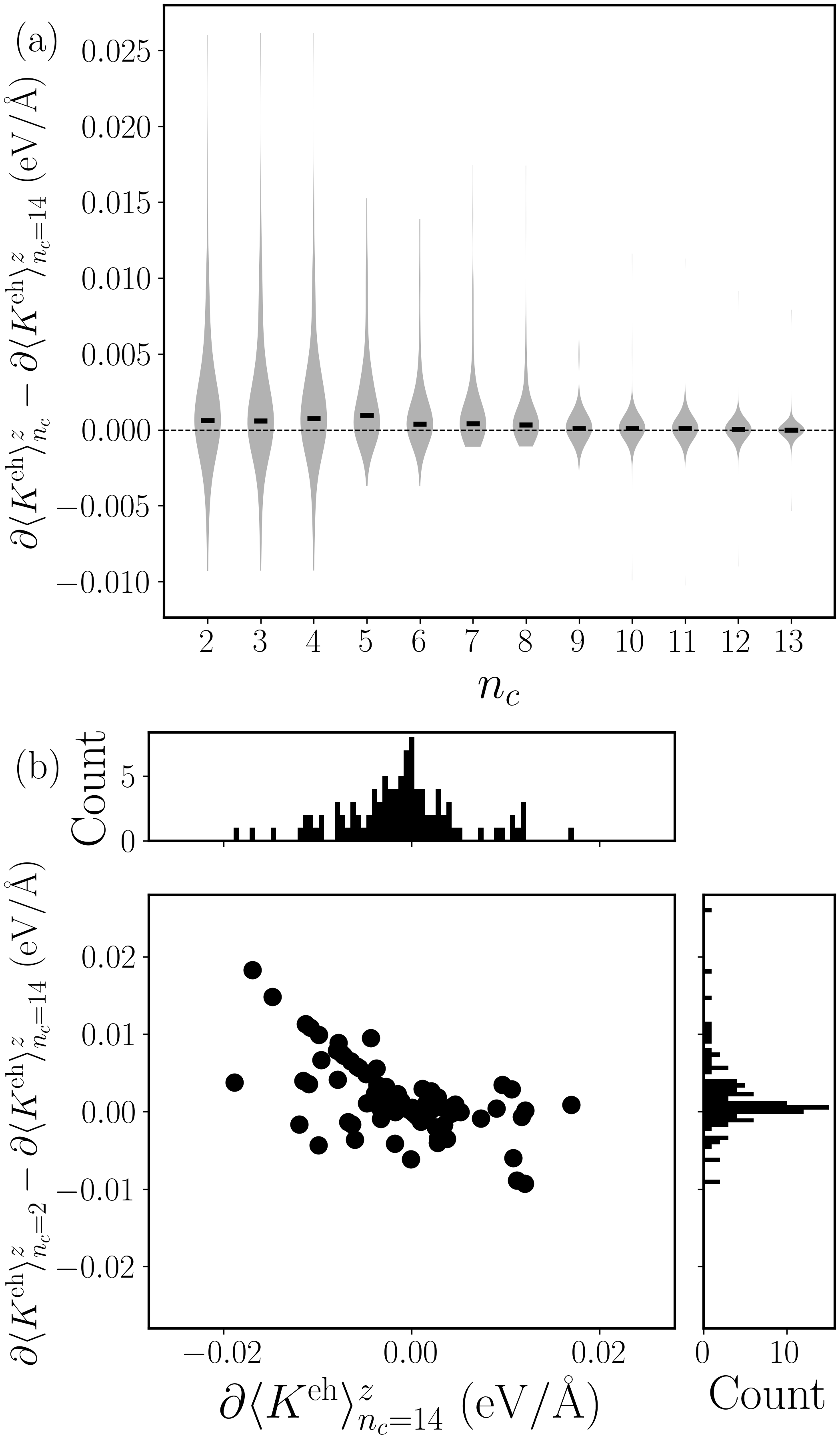}
    \caption{(a) Evolution of the distribution of  $\partial \langle K^{\rm{eh}} \rangle_{n_c} - \partial \langle K^{\rm{eh}} \rangle_{n_c=14}$, using equation \ref{eq:expansion_kernel_derivatives_ibl}, as function of the number of conduction states $n_c$ used in the expansion sums. (b) Comparison between the difference $\partial \langle K^{\rm{eh}} \rangle_{n_c} - \partial \langle K^{\rm{eh}} \rangle_{n_c=14}$ and $\partial \langle K^{\rm{eh}} \rangle_{n_c=14}$, which are about the same order, showing the importance of including all terms to obtain converged forces.}
    \label{fig:convergence_der_kernel_partial_sum}
\end{figure}

Then we move to the analysis of the convergence of extra sums used in equation \ref{eq:expansion_kernel_derivatives_ibl} for solids. For simplification we denote here $\langle K^{\rm{eh}}\rangle = \sum_{\mathbf{k}cv \mathbf{k}'c'v'}  \langle \mathbf{k}cv|K^{\rm{eh}}|\mathbf{k}'c'v'\rangle$. Differently than in the molecules' case, the convergence of the BSE in solids is achieved with a relative small number of conduction and valence states. We solve the BSE for monolayer MoS$_2$ with 14 conduction bands and calculate the sums in equation \ref{eq:expansion_kernel_derivatives_ibl} from equation by varying the number conduction bands used in them from 2 to 14 for the first one-hundred excitons, and we analyze the $z$ (out of plane) component of this force over one of the sulfur atoms. In figure \ref{fig:convergence_der_kernel_partial_sum}-a we see that the dispersion of $\partial \langle K^{\rm{eh}} \rangle_{n_c} - \partial \langle K^{\rm{eh}} \rangle_{n_c=14}$ decreases monotonically to our reference calculation with variations up to 0.025 eV/\text{\AA}. In figure \ref{fig:convergence_der_kernel_partial_sum}-b we look in detail the worst case ($n_c=2$), and observe that the variation of $\partial \langle K^{\rm{eh}} \rangle_{n_c} - \partial \langle K^{\rm{eh}} \rangle_{n_c=14}$ is in the same order of $\partial \langle K^{\rm{eh}} \rangle_{n_c=14}$, showing that summations in equation \ref{eq:expansion_kernel_derivatives_ibl} need to be done will all bands used to build the BSE Hamiltonian. In this analysis we used 100 excitons, which include high-energy excitons up to 3.1 eV. When we use 10 excitons the maximum errors decrease to up to $\sim$ 0.003 eV/\text{\AA}, as low-energy excitons in MoS$_2$ need less states to be converged. 

\section{Excited state forces in LiF}
\label{sec:benchmark_LiF}

Now we move to the analysis of ESF on LiF. We start computing ESF for its rocksalt structure with $O_h$ point group symmetry. We observe that the excited state forces are essentially zero due to the system symmetry. One can apply displacements to break the system symmetry to relax the excited state. Those displacements can be random or based on physical intuition about a particular case. For our demonstration, we applied random displacements using gaussian distribution with a standard deviation temperature dependent equal to $\sigma_i = \sqrt{k_b T / \lambda_i}$, where $\lambda_i$ is the eigenvalue $i$ of the force constant matrix $k$ obtained from DFPT. Once the random displacement is applied then we relax the system using Newton's method. Other methods as steepest descent and conjugate gradient are also perfectly good options (see section \ref{sec:relaxation_schemes}). For each relaxation step, we calculate the analytical excited state force and force constant matrix from DFPT. The total displacement is given by

\begin{equation}
    \delta r = \sum_i \hat{\lambda}_i \mathrm{min} ( (\vec{F} \cdot \hat{u}_i ) / \lambda_i, \mathrm{limit})
\end{equation}

\noindent where $\lambda_i$ and $\vec{\lambda}_i$ are the $i$-th eigenvalue and the $i$-th eigenvector of $k$.

\begin{figure}[ht]
\includegraphics[width=\columnwidth]{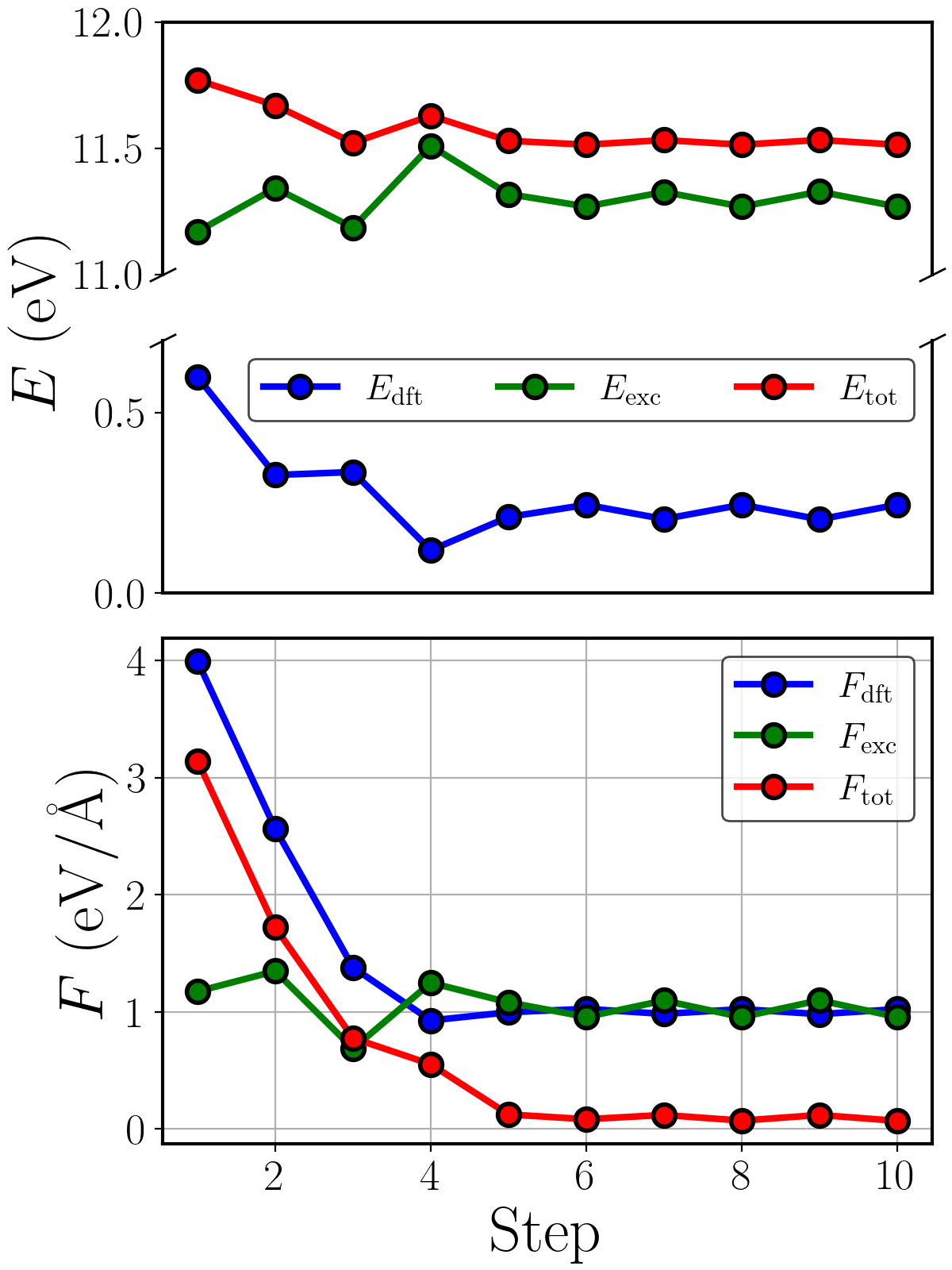}
\caption{Energies variations (upper panel) and forces (lower panel) on LiF during relaxation of the ESF summed to the DFT force. Step zero corresponds to the RS structure, for which both the ESF and DFT forces are null due to symmetry. Step one corresponds to random displacement and step ten corresponds to the STE configuration. }
\centering
\label{fig:LiF_1unitcell_relaxation_Newton_method}
\end{figure}

The results for our relaxation are shown in figure \ref{fig:LiF_1unitcell_relaxation_Newton_method}. We observe that the total energy decreases until convergence. We also monitor the total force that reaches zero. 

This shows how our analytical ESF can be used to study self-trapped excitons in materials. This study for LiF corresponds to 1 exciton per unit cell (6$\times$10$^{22}$ excitons / cm$^3$), which is an extreme exciton concentration. The exciton trapping energy is about 150 meV and the redshift in the absorption spectra is 0.4 eV. The relaxed structure belongs to $C_{3v}$ point group while the rocksalt LiF belongs to $O_h$.  We observe the valence band at the $\Gamma$ point, originally triple degenerate, splits in one double degenerate and one single band and there is a decrease of the main gap at the $\Gamma$ point. This leads to the splitting of the first absorption peak, creating one redshifted and one blueshifted peak.

\subsection{Symmetry-based path of trapping excitons}

\begin{figure}[ht]
    \centering
    \includegraphics[width=\columnwidth]{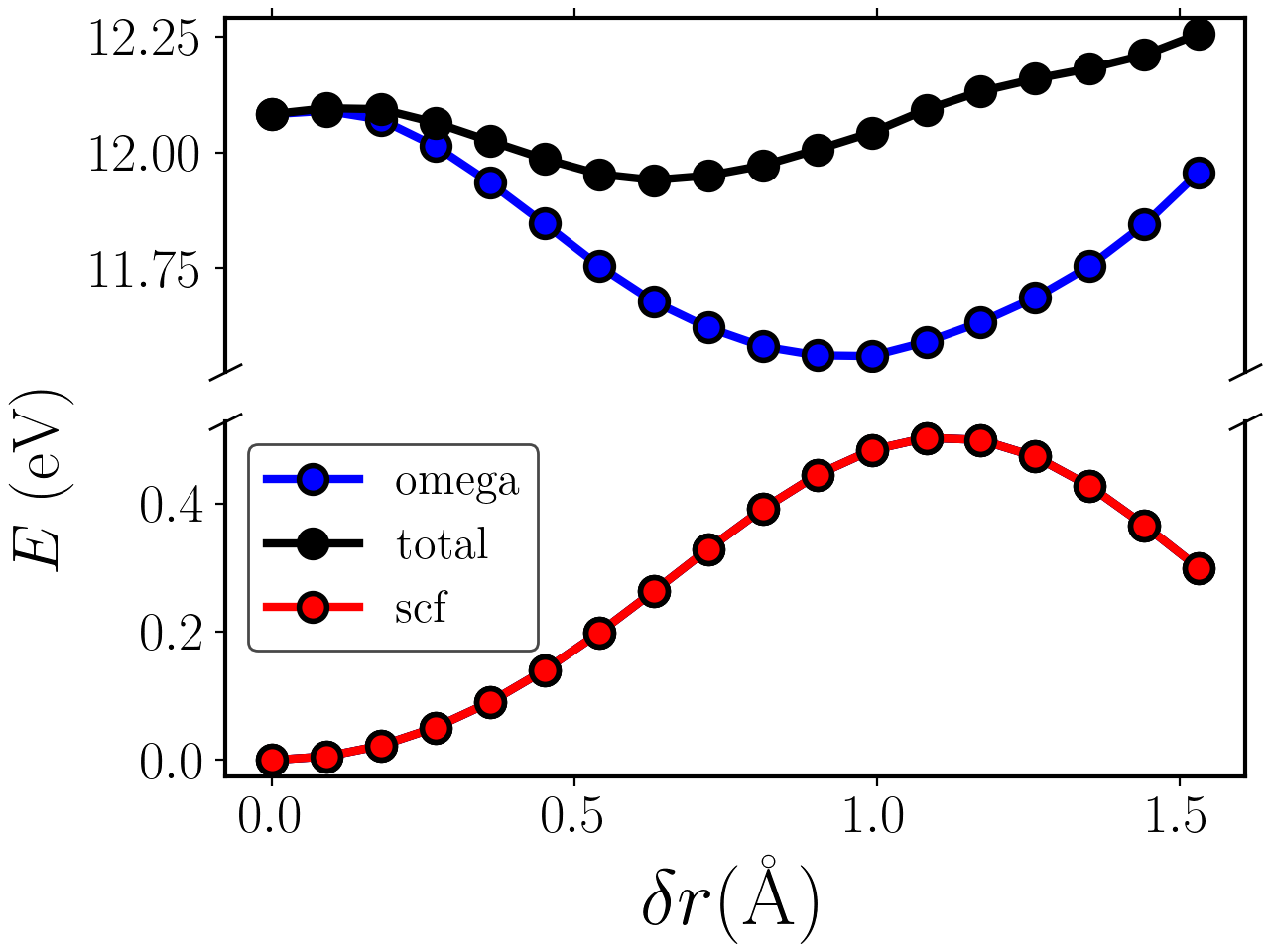}
    \caption{DFT, excitation and total energies for the first exciton as function of the displacement of the F atom through LiF unit cell. The $C_{3v}$ symmetry is preserved during this displacement.}
    \label{fig:LiF_energies_path_C_3v}
\end{figure}

As the relaxed excited state configuration has $C_{3v}$ symmetry we now investigate the transition from rocksalt to this relaxed structure. To do this we move the fluoride atom in the direction where the $C_{3v}$ symmetry is preserved (see Fig. \ref{fig:LiF_energies_path_C_3v}). We see that initially the DFT increases and the first exciton energy decreases, while the total energy has a minimum at $\delta r \approx 0.6 $ \text{\AA}. We observe a small energy barrier of 13 meV. 

\begin{figure}[ht]
    \centering
    \includegraphics[width=\columnwidth]{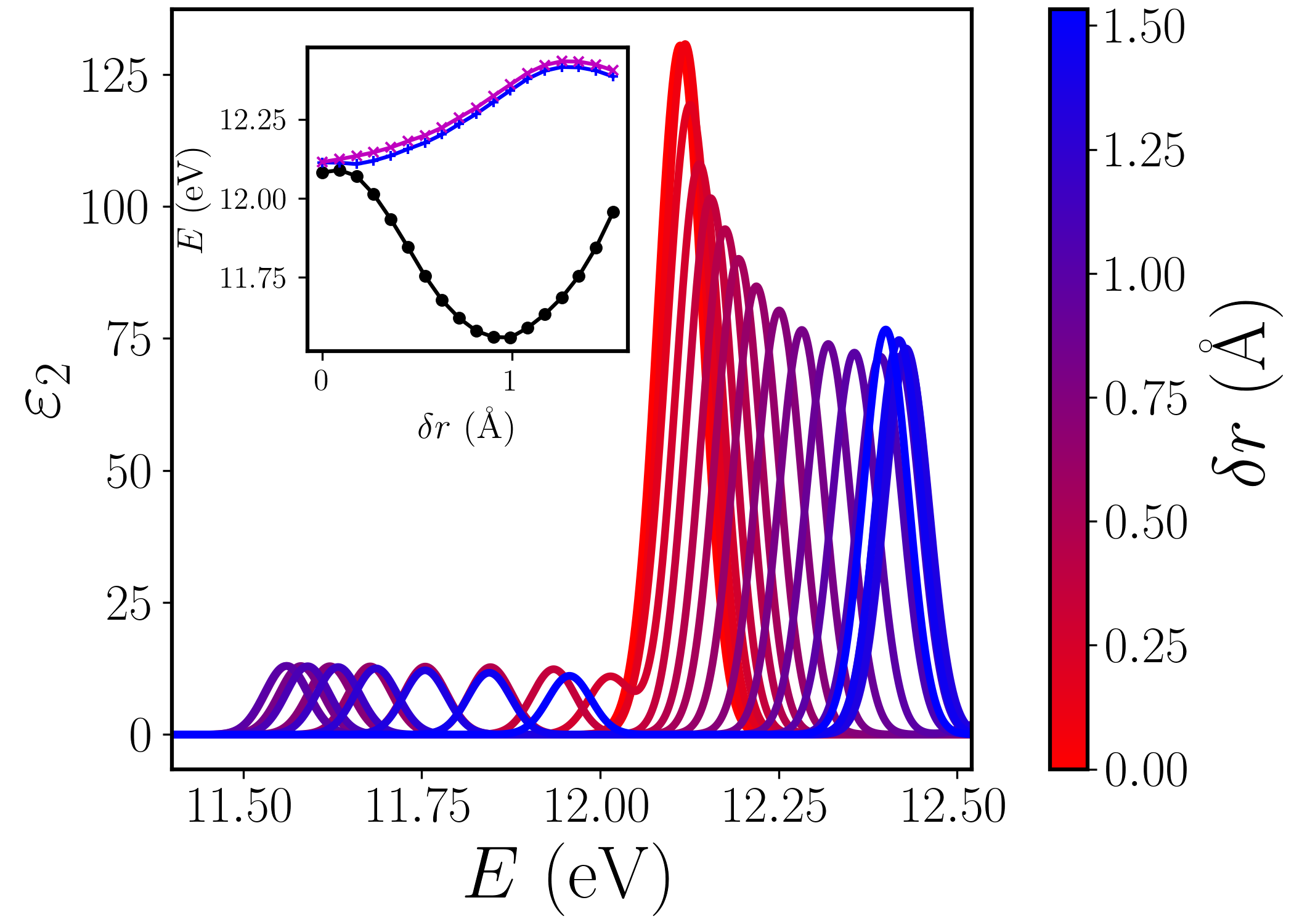}
    \caption{Evolution of the absorption spectra of LiF. The main peak at 12.1 eV splits in one blueshifted and one redshifted  peaks. The energies of the first three excitons are also shown in the inset of the figure.}
    \label{fig:evolution_abs_spectra_LiF}
\end{figure}

In Figure \ref{fig:evolution_abs_spectra_LiF} we show the evolution of the LiF absorption spectra as we move the fluorine atom. The peak initially at 12.1 eV and composed by three excitons, splits in two new peaks. One peak is composed by two excitons and is blueshifted while other is composed by one exciton is initially redshifted and then after $\delta r \approx 0.6 $ \text{\AA} this  peak is blueshifted. We also show the evolution of the energies of the three first excitons.

\begin{figure}[ht]
    \centering
    \includegraphics[width=\columnwidth]{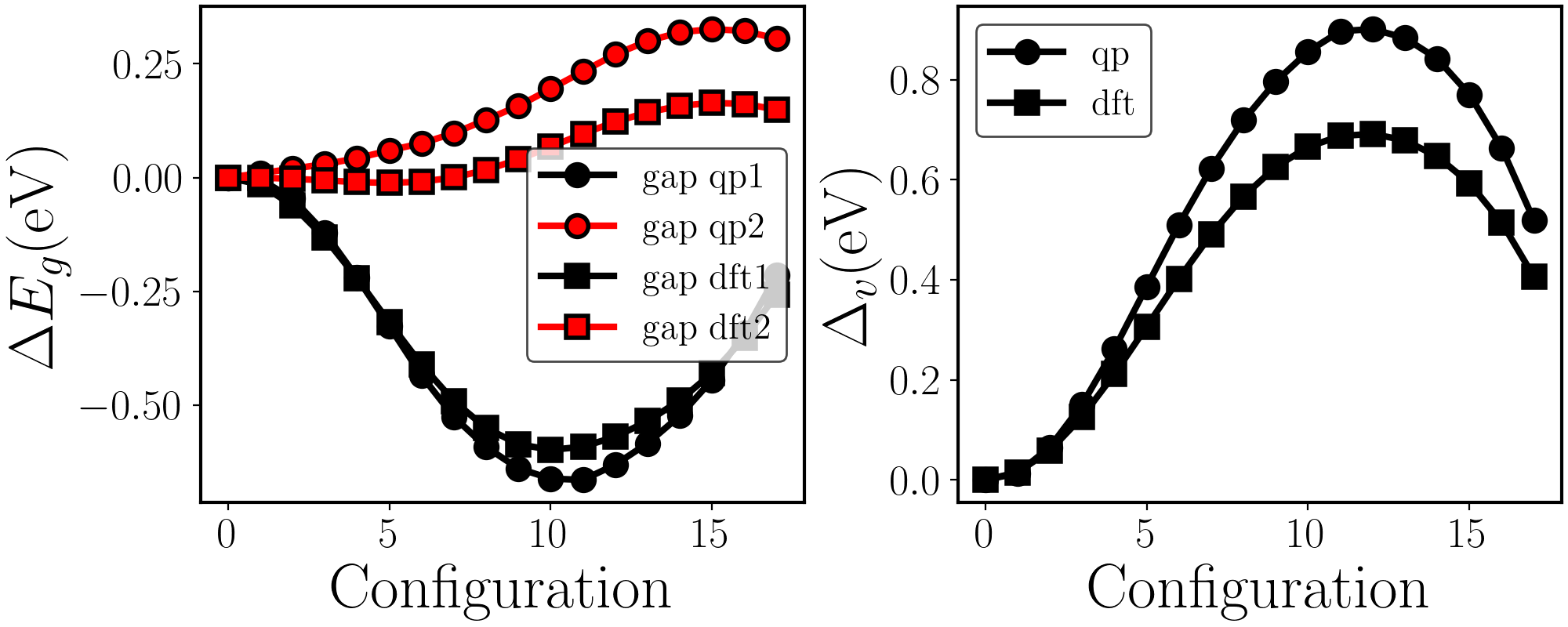}
    \caption{Left: First and second gaps at the $\Gamma$ point for LiF as function of displacement of F atom. Right: split of the valence band.}
    \label{fig:LiF_change_of_gap_and_split_of_valence_energy_levels}
\end{figure}

The split of the triply degenerate exciton is explained by its bandstructure. At the $\Gamma$ point the valence band is triply degenerate and has $p$ orbital character , while the conduction band is non-degenerate. As we move the fluoride atom this triply valence band breaks in one doubly degenerate and one non-degenerate. In Figure \ref{fig:LiF_change_of_gap_and_split_of_valence_energy_levels} we show the split of the valence band. This leads to two gaps close in energy, and one gap increases and the other decreases as we move the fluoride atom.

\begin{figure}[ht]
    \centering
    \includegraphics[width=\columnwidth]{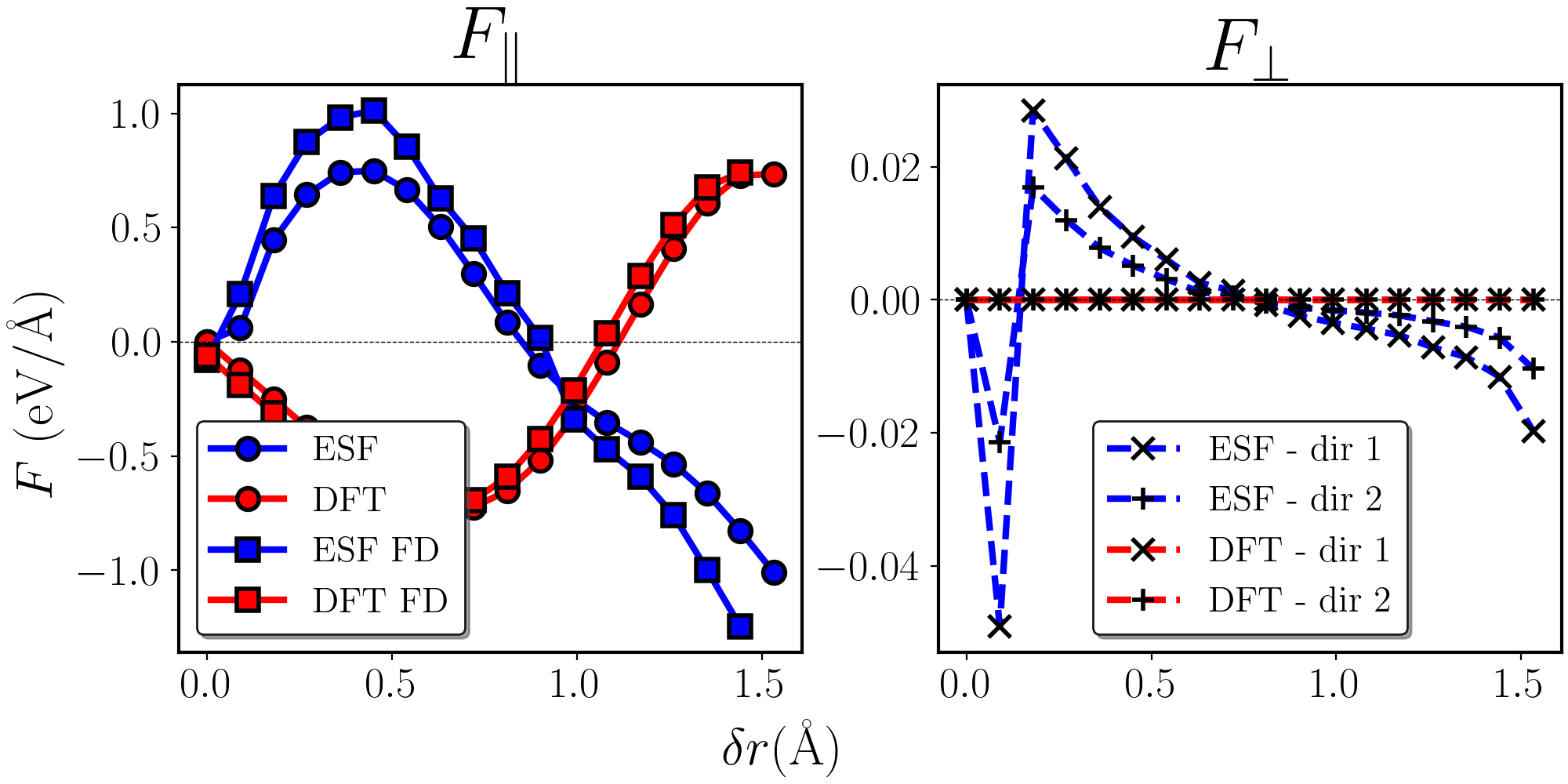}
    \caption{ESF and DFT forces on the F atom as function of its displacement. On the left we show the forces components parallel to the atomic displacements, while in the right we show components perpendicular. The parallel component is much larger than the perpendicular components.}
    \label{fig:LiF_forces_path_C_3v-label}
\end{figure}

Along this path, both the excited state force and the DFT force are parallel to the atomic displacement as shown in figure \ref{fig:LiF_forces_path_C_3v-label}. We obtain a good agreement between analytical ESF and FD ESF, which shows the efficiency of our method.

\subsection{Derivatives of elph matrix elements - STE}

\begin{figure}[ht]
\includegraphics[width=\columnwidth]{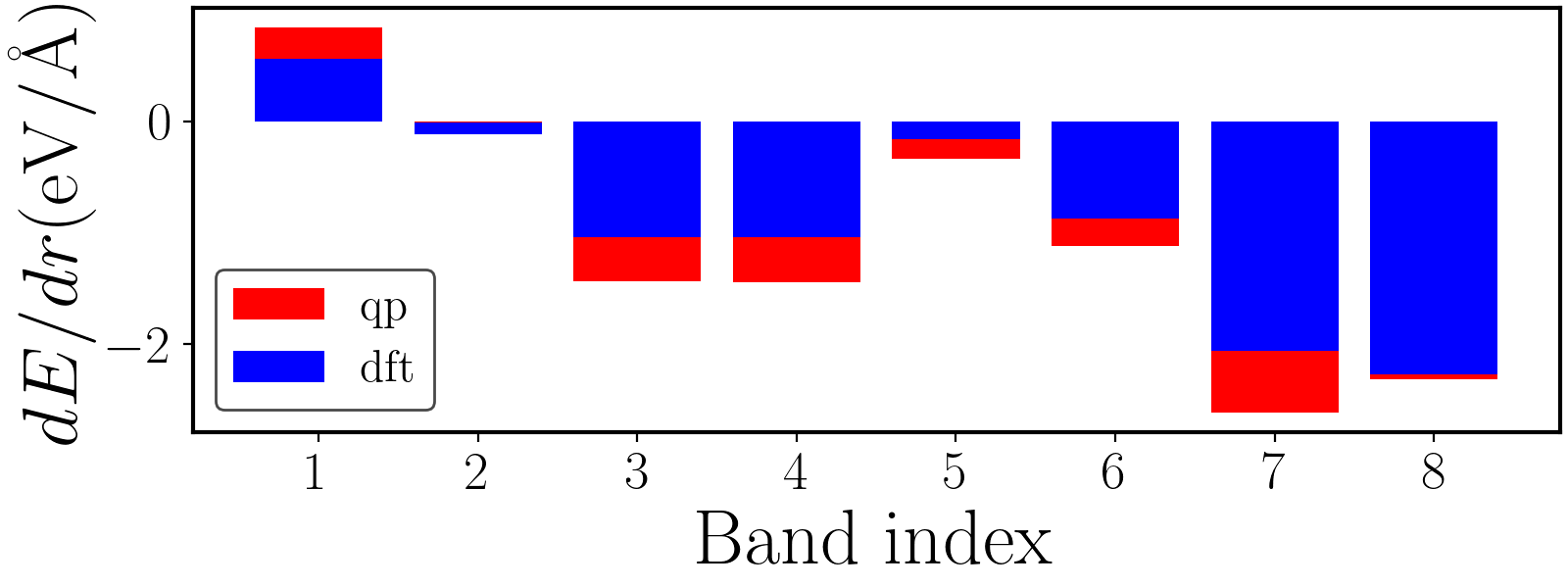}
\caption{Derivative of DFT (blue bars) and QP (red bars) energy levels for the first eight bands in LiF at the $\Gamma$ point. The first five bands are valence bands.}
\centering
\label{fig:derivative_QP_DFT_energy_levels_LiF}
\end{figure}

We now analyze the performance of our approximations on ELPH coefficients and kernel derivatives for LiF in the STE configuration, as in the rocksalt configuration diagonal ELPH coefficients and kernel derivatives are null due to symmetry. First, we look at the diagonal ELPH coefficients, which are equal to the derivatives of QP and DFT energy levels. We observe that in general the DFT derivatives underestimate QP derivatives about 30\% for the eight first bands in LiF for all $k$ points in LiF first BZ. In figure \ref{fig:derivative_QP_DFT_energy_levels_LiF} we show the derivatives of energy levels at $\Gamma$. 

\begin{figure}
    \centering
    \includegraphics[width=\columnwidth]{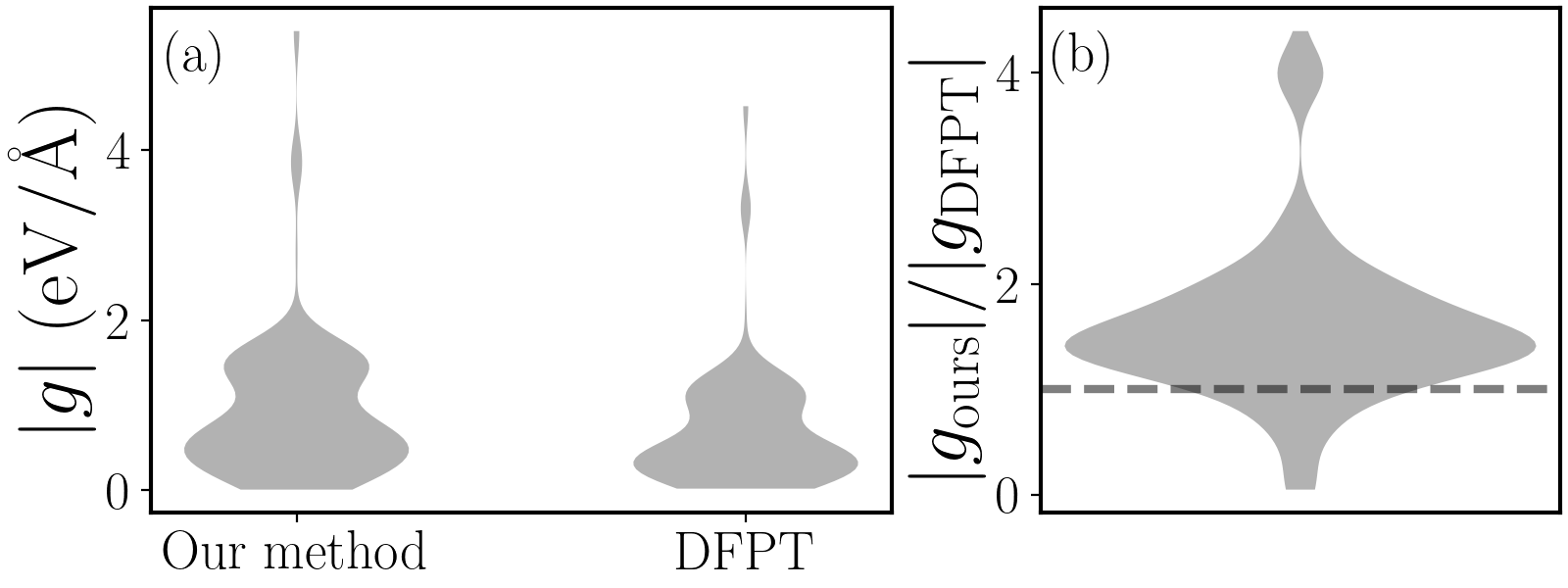}
    \caption{(a) Distribution of ELPH coefficients using our renormalization method and raw DFPT. (b) Ratio between renormalized and raw ELPH coefficients. Here we analyze the three first valence and the first conduction bands for LiF in a 4$\times$4$\times$4 k-grid.}
    \label{fig:elph_violin_plots_LiF}
\end{figure}

Then we analyze the off-diagonal ELPH matrix elements. In general, our renormalization scheme increases the magnitude of ELPH coefficients, as shown in figure \ref{fig:elph_violin_plots_LiF}. 



\subsection{Changes in Kernel - STE}

\begin{figure}[ht]
\includegraphics[width=\columnwidth]{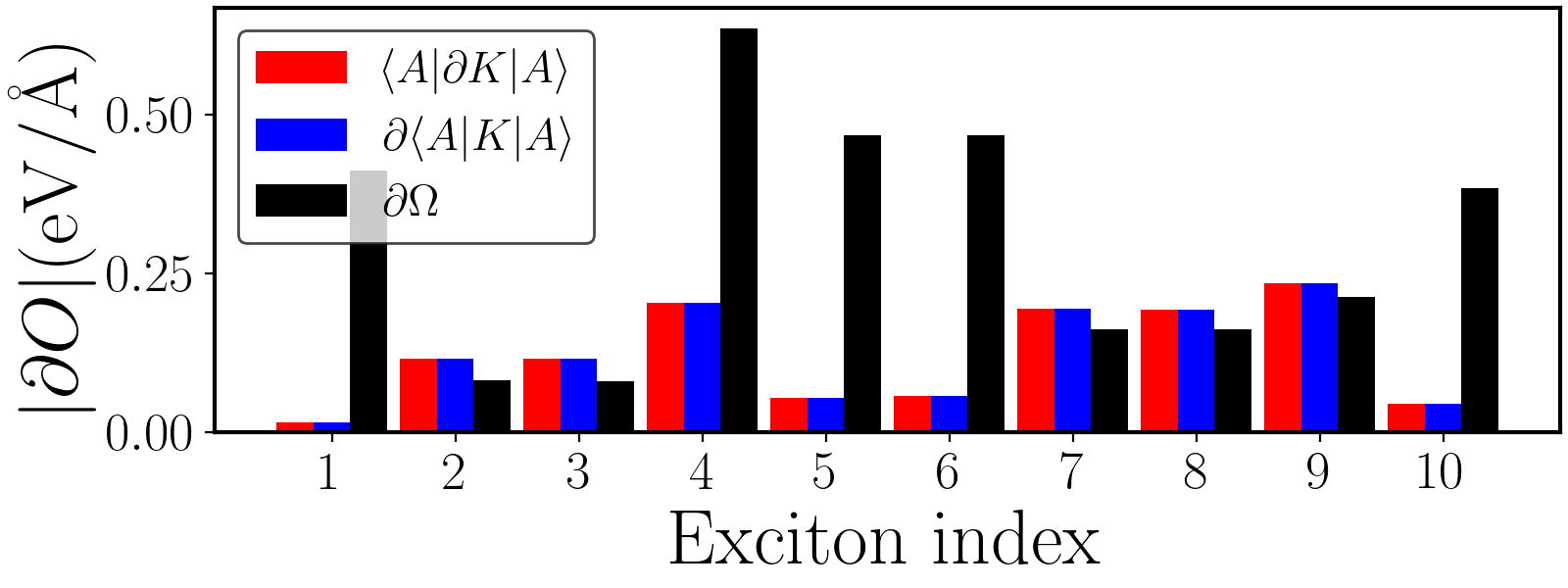}
\caption{Derivatives of kernel compared to derivatives of the excitation energy for LiF.}
\centering
\label{fig:LiF_Derivative_kernel_vs_derivative_omega}
\end{figure}

Now we look at the derivatives of kernel. On figure \ref{fig:LiF_Derivative_kernel_vs_derivative_omega} we observe that in general derivatives of kernel are a fraction of the derivatives of the ESF, although in cases where $\partial \Omega$ is smaller than 0.25 eV/\text{\AA} the derivatives of the kernel are of the same order as the derivatives of the excitation energy.

%
%

\section{Relaxation schemes}
\label{sec:relaxation_schemes}

\begin{figure}
    \centering
    \includegraphics[width=\linewidth]{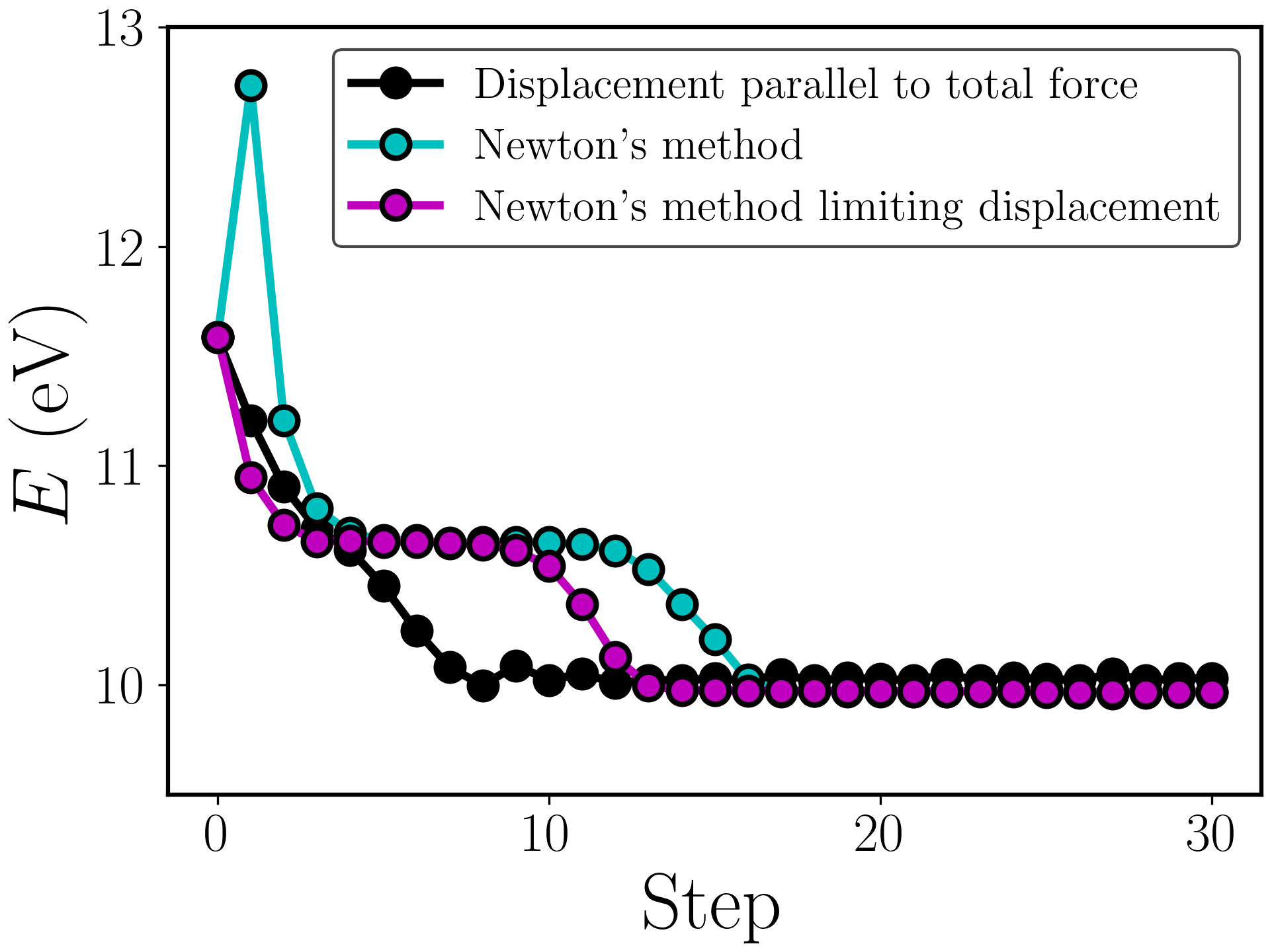}
    \caption{Comparison of Newton's method and our steepest descent implementation in the relaxation process of the excited state of a $1\times 1 \times 2$ LiF supercell (see text). }
    \label{fig:comparison_relaxation_schemes}
\end{figure}

In this section we highlight some strategies to perform relaxations of excited states. Schemes such steepest descent and conjugate gradient are useful options, although the high computational cost of GW/BSE and DFPT calculations compared to simple DFT self-consistent field iterations limits the use of ESF to simulations with relative small number of atoms. Regarding this we compare different strategies of relaxation to a 1$\times$1$\times$2 LiF supercell. 
\begin{itemize}
    \item Our first strategy is to make use of our approximation of constant excited state force. We also approximate that the force constants for the excited state are the same of the ground state (i.e. $\partial^2 \Omega / \partial r_\alpha \partial r_\beta \approx 0$) and then we can use the force constant matrix $k$ obtained from DFPT. Using this information one can look for a displacement parallel to the total force $\delta \vec{d}=\alpha\vec{F}=\alpha(\vec{F}_{ex}+\vec{F}_{DFT})$ and find the optimal $\alpha$ that minimizes the following harmonic approximation to the total energy
\begin{equation}
    \Delta E\approx -\delta \vec{d} \cdot (\vec{F}_{ex}+\vec{F}_{DFT}) + \frac{1}{2} \delta \vec{d}^T k \delta \vec{d}
\end{equation}
By projecting the total force in the basis of eigenvectors $\vec{\lambda}$ and eigenvalues $\lambda$ of the force constant matrix, one can obtain the optimum proportionality constant is 
\begin{equation}
    \alpha = \frac{\sum_\lambda F^2_\lambda}{\sum_\lambda \lambda F^2_\lambda}
\end{equation}
where $F_\lambda = \vec{F} \cdot \vec{\lambda}$ . 
\item In the same spirit of the previours strategy, we expand both the displacement $\delta \vec{d}=\sum_\lambda d_\lambda \vec{\lambda}$ and the total force $\vec{F} = \sum_\lambda F_{\lambda} \vec{\lambda}$ in eigenvectors of $k$ and obtain that each displacement component is given by $d_\lambda=F_\lambda / \lambda$ (for $\lambda \neq 0$). For the acoustic modes we set $F_\lambda =0 $, so the total force in the system center of mass is zero. This approach is also refereed as Newton's method, but here force constant are obtained from finite differences instead of DFPT calculations. 
\item Using the approximation that the ESF is constant, one can use those forces as constant external forces in DFT relaxations. This is possible to be done with Quantum Espresso \cite{Giannozzi2009, Giannozzi2017, Giannozzi2020} and many other DFT codes, and it goes beyond the harmonic approximation for the ground state.
\end{itemize}
After applying the desired displacement, one needs to update the excited state forces until a convergence criteria is achieved. In order to save computational time, a possible strategy is to update the just the ELPH coefficients or just the $A_{\mathbf{k}cv}$ coefficients (by performing again GW/BSE calculations). In this approach the complex phase information is lost, then one needs to use just diagonal ELPH matrix elements for the ESF expression, leading to $\vec{F} = \sum_{\nu \mathbf{k}cv} |A_{\mathbf{k}cv}|^2(g^\nu_{\mathbf{k}c,\mathbf{k}c}-g^\nu_{\mathbf{k}v,\mathbf{k}v}) \hat{u}^\nu$. In the case of defects, where excitons are composed by a single $\mathbf{k}v \to \mathbf{k}c$ transition ($A_{\mathbf{k}cv} = \delta_{\mathbf{kk'}} \delta_{cc'} \delta_{vv'}$) the ESF expression reduces to $\vec{F}=-\sum_\nu  (g^\nu_{\mathbf{k}c, \mathbf{kc}}-g^\nu_{\mathbf{k}v, \mathbf{k}v}) \hat{u}^\nu$, and this strategy is similar to constrained DFT used in reference \cite{Kirchhoff2024}.  GW/BSE calculations can be speed-up with the use of scissor operator \cite{Deslippe2012} or reusing the dielectric function from the initial step into the next steps. The validity and accuracy of those approximations are out of the scope of this work and will be studied in the future. In figure \ref{fig:comparison_relaxation_schemes} we compare our steepest descent with Newton's method implementation to relax the excited state in a $1\times 1\times 2$ supercell of LiF. In general the steepest descent worked better for the relaxation. 

\section{Polaronic excitons in LiF}

\label{sec:polaronic_exc_lif}

\begin{figure}
    \centering
    \includegraphics[width=\linewidth]{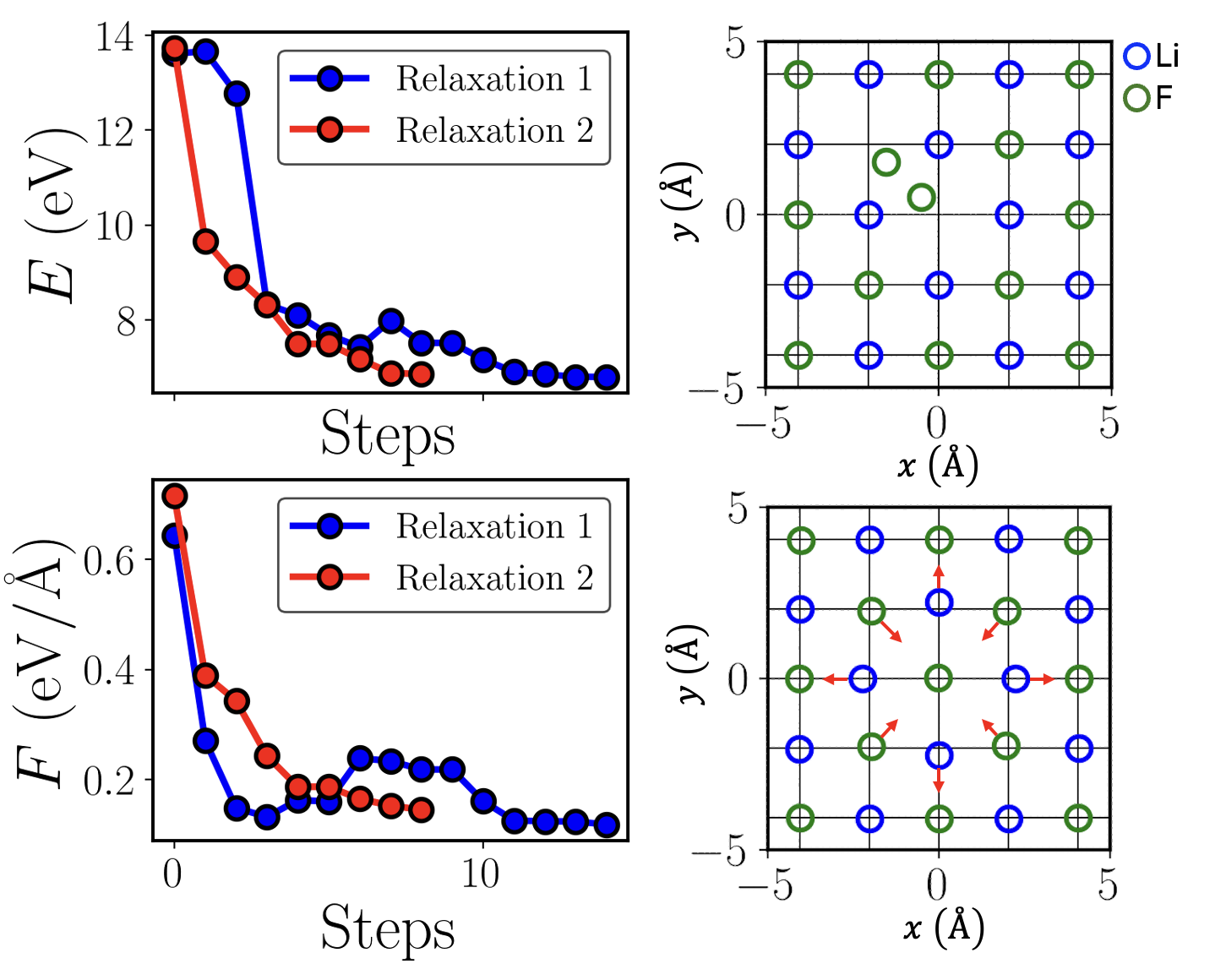}
    \caption{Convergence of total energy (a) and mean force (b) in the relaxation processes of the polaronic exciton in LiF. Initial (c) and final (d) atomic positions of atoms in the (100) plane for the first relaxation. The initial displacement consists of two fluorine atoms moved towards each other \cite{delgrande2025}. In the final configuration the polaronic exciton is centered in a fluorine atom, where the first Li neighbor atoms move away from it and fluorine atoms move towards it (see red arrows in (c))}
    \label{fig:lif_sc_exc_relaxation}
\end{figure}

\begin{figure}
    \centering
    \includegraphics[width=\linewidth]{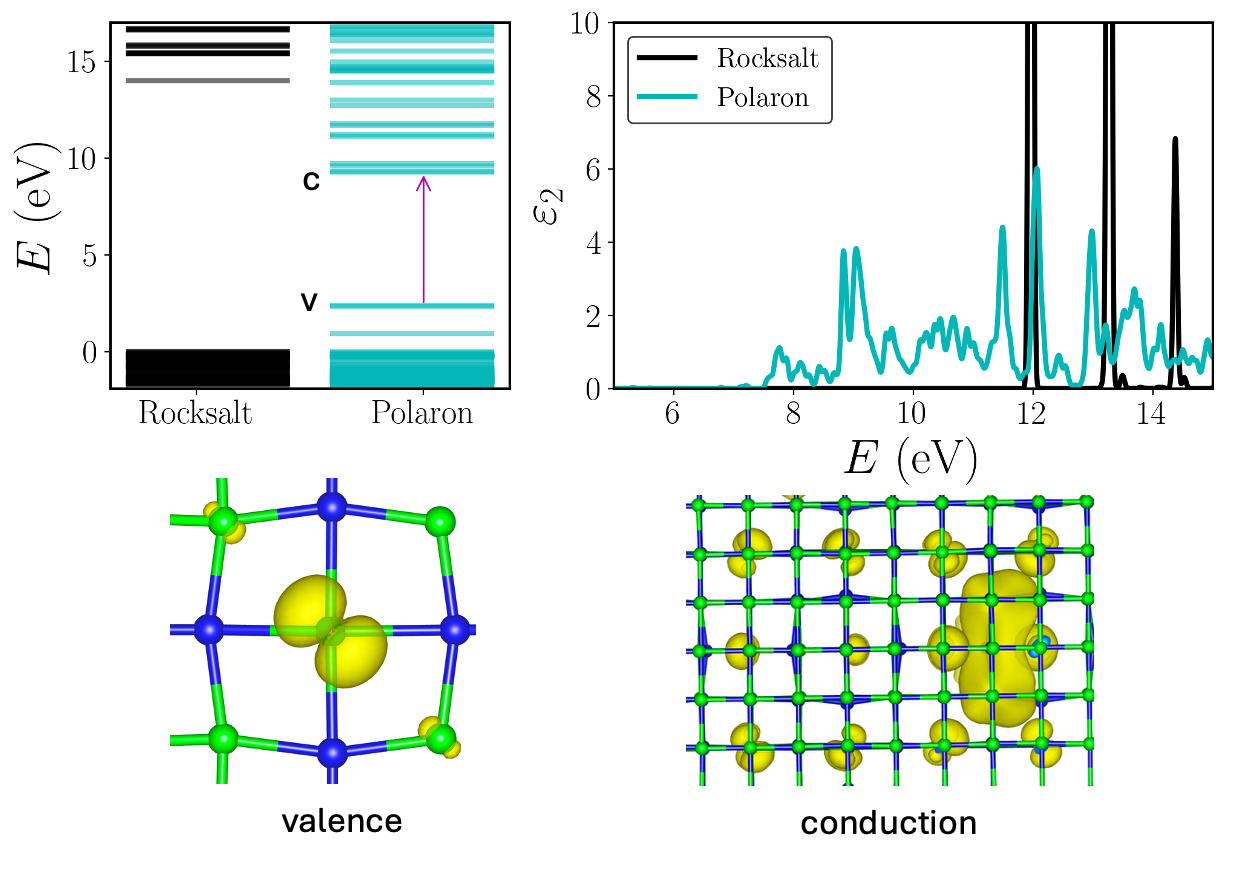}
    \caption{(a) Energy levels of LiF in a $4 \times 4\times 4$ supercell at the rocksalt and polaronic exciton configurations. Vertical arrow represents the first excited state, whose total energy was relaxed (see figure \ref{fig:lif_sc_exc_relaxation}. (b) Optical absorption of rocksalt and polaronic exciton configurations. (c) and (d) are the valence and conduction states, respectively, that compose the polaronic exciton.}
    \label{fig:electronic_properties_polaronic_exc}
\end{figure}

In this section we apply the excited state forces to LiF in a 4$\times$4$\times$4 supercell. To save computational time we choose the number of valence and conduction bands to be 137 and 31, respectively by using our method to build BSE Hamiltonian with a reasonable size, where we just use 12\% of the matrix elements of the zone-folding Hamiltonian with a accuracy of 0.15 eV for the lowest-energy exciton binding energy \cite{delgrande2025}. In the rocksalt arrangement the ESF over the Li and F atoms is zero by symmetry, therefore before the relaxation we need to apply atomic displacements that break the symmetry. We obtained polaronic excitons similar to those obtained by Dai et. al \cite{Dai2024PRL, Dai2024PRB} in two  relaxations with two different starting points. In the first relaxation we applied displacements similar to the ones described in \cite{Dai2024PRL, Dai2024PRB}, while in the second relaxation our starting point is the rocksalt structure with two fluorine atoms moved towards each other, trying to mimic a V$_k$ center (see figure \ref{fig:lif_sc_exc_relaxation}), similar to the test we did in reference \cite{delgrande2025}. We expected the second relaxation to converge to a V$_k$ center configuration, although it converged to the polaronic exciton configuration as our first relaxation did. One may expect that to simulate V$_k$ center it is necessary to work with open-shell systems, which is out of the scope of this work. The fact of both independent relaxations converged to the same polaronic exciton show the accuracy of our ESF implementation. Our final displacements of the polaronic exciton configuration in relation to the rock agree with the patterns obtained in reference \cite{Dai2024PRB}. Our polaronic exciton is composed by a localized hole state and a delocalized electron state (see figure \ref{fig:electronic_properties_polaronic_exc}). 

We presented self-trapped exciton results in LiF for different types of calculations, showing the diversity of self-trapped excitons (STE) that may occur in LiF. On table \ref{tab:optical_properties_polaronic_exc_lif} we report the Stokes shift and Zero Phonon Line (ZPL) for those excitons. 

\begin{table}[h]
    \centering
    \begin{tabular}{|c|c|c|l|}
     \hline
                 & 1 PC & 2 PC  &64 PC\\
        \hline
        $\Omega_{\rm{STE}}$ & 11.7 &  7.9   &5.3\\ 
        \hline
        ZPL & 11.9 & 10.0 &6.8\\
        \hline
        Stokes shift & 0.4 & 4.2  &6.6\\
        \hline
        
    \end{tabular}
    \caption{Exciton energy for LiF in rocksalt structure is 12.08 eV. $\Omega_{\rm{STE}}$ is the excitation energy of the self-trapped exciton ($\Omega_{\rm{STE}}$). We also report calculated Zero-phonon line (ZPL) and Stokes shift ($=E_{\rm{rocksalt}}-E_{\rm{STE}})$) for calculations using only one primitive cell (1 PC), two primitive cells   (2PC) and a 4$\times$4$\times$4 (64 PC) supercell that converged to a polaronic exciton shown in figures \ref{fig:lif_sc_exc_relaxation} and \ref{fig:electronic_properties_polaronic_exc}.}
    \label{tab:optical_properties_polaronic_exc_lif}
\end{table}

\section{ESF on phonon basis and exciton-phonon symmetries}

\label{sec:exc_ph_symm}

\begin{figure}
    \centering
    \includegraphics[width=\linewidth]{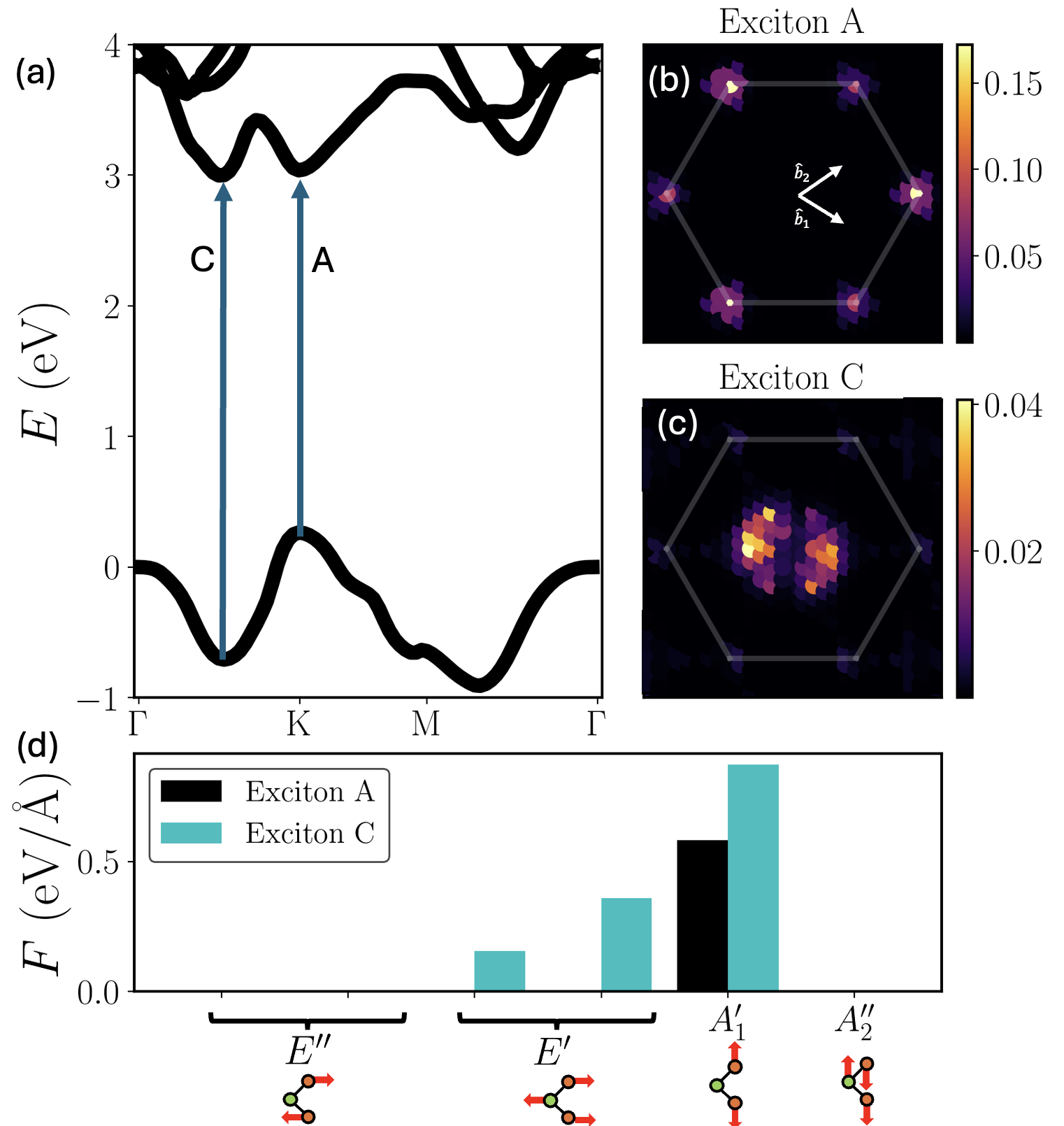}
    \caption{(a) Bandstructure of monolayer MoS$_2$ at GW level. Vertical arrows represent transitions at $K$ and $\alpha K$ ($0< \alpha < 1$) that composes the $A$ and $C$ excitons, respectively. (b) and (c) are the distributions of $M(\mathbf{k}) = \sum_{cv} |A_{\mathbf{k}cv}|^2$ over the first Brillouin Zone of MoS$_2$ for exciton $A$ and $C$, respectively. In (d) we show the projection of the excited state forces for excitons $A$ and $C$ on the phonon displacement of monolayer MoS$_2$.}
    \label{fig:forces_ph_1l_mos2}
\end{figure}

The ESF can be projected in the phonon displacement basis for an analysis of exciton-phonon interactions. In reference \cite{Karkee2026AdvMatTech}, we found that the lowest energy excitons in the 1D perovskite C$_4$N$_2$H$_14$PbBr$_4$ coupled with its 100 cm$^{-1}$ phonon mode, in agreement with Huang-Rhys analysis of the experimental data of temperature dependent photoluminescence peak width from \cite{Yuan2017}. In figure \ref{fig:forces_ph_1l_mos2}, we project the excited state forces for the A and C excitons of monolayer MoS$_2$ in its phonon displacement basis. The exciton A couples with the $A'_1$ phonon mode, while exciton C couples with the $A'_1$ and $E'$ phonon modes. Transient absorption experiments in 1L MoS$_2$ shows that the C exciton couples with the $A'_1$ phonon mode \cite{Trovatello2020}, while resonant Raman experiments show that $A'_1$ phonon mode is enhanced by A and B excitons and $E'$ phonon mode is enhanced by C excitons \cite{CarvalhoPRL2015}. 

Our results and these experimental observations can be explained by group theory analysis of symmetries \cite{dresselhaus_book, nalabothula2025symmetriesexcitons}. Here we focus on the A and C exciton, as the B exciton is similar to A exciton, with difference between both being due to spin orbit effects \cite{Qiu2016}. Monolayer MoS$_2$ belongs to the $D_{3h}$ point group. The A exciton is composed by transitions at $K$ point (see figure \ref{fig:forces_ph_1l_mos2}), which little group is $C_{3v}$, while the C exciton is composed by transitions for $k$-points around $\Gamma$ \cite{QiuPRL2013}, then we analyze it as a transition at a point between $\Gamma$ and $K$, which little group is $C_s$. For an exciton-phonon matrix element $\langle A| \partial_\nu H^{\rm{BSE}}| A \rangle$ to be non zero it is necessary that the direct product $D(\Psi_{\rm{exc}})\otimes D(\nu)\otimes D(\Psi_{\rm{exc}})$ contains the the identity representation \cite{dresselhaus_book}, where $D(\Psi_{\rm{exc}})$ is the irreducible representation of the exciton and $D(\nu)$ is the irreducible representation of the phonon $\nu$. We consider those two excitons to be Wannier excitons as their radius are about a few nm \cite{QiuPRL2013}. The representation of a Wannier exciton composed by a transition from a valence band $v$ to a conduction band $c$ is given by $D(\Psi_{\rm{exc}}) = D(\psi_c)\otimes D(\psi_v) \otimes D(F)$, where $F$ is the envelope function for this Wannier exciton given by $\Psi_{\rm{exc}}(\mathbf{r}_r, \mathbf{r}_h) = F(\mathbf{r}_e-\mathbf{r}_h ) \psi_{\mathbf{k}c} (\mathbf{r}_e) \psi^*_{\mathbf{k}v} (\mathbf{r}_h)$. Both the A and C exciton are the lowest states (1s) of their respective series, so their envelope functions transform as the totally symmetric representation and the exciton representation reduces to $D(\Psi_{\rm{exc}})=D(\psi_c) \otimes D(\psi_v)$. In table \ref{tab:group_theory_analysis} we report the irreducible representations of the $A$ and $C$ excitons and the conduction and valence states that compose those excitons. Therefore, for both excitons, group theory analysis predicts that both $A'_1$ and $E'$ phonon modes couple with the $A$ and $C$ excitons, in agreement with our calculations and experimental data \cite{CarvalhoPRL2015, Trovatello2020}, with exception of the coupling between $E'$ phonon mode with the $A$ exciton, as group theory predicts if a matrix element is non-zero but it does not state its value. 

\begin{table*}
    \centering
    \begin{tabular}{|c|c|l|c|c|c|c|}
    \hline
         Exciton & Transition at  &Little group& $D(\psi_c)$ & $D(\psi_v)$ & $D(\Psi_{\rm{exc}})$ & Phonons that couple with this exciton  \\
     \hline
         A & $K$ &$C_{3h}$& $A'$ & $E'$  & $E'$ & $A'_1$, $E'$  \\\hline
         C & Between $\Gamma$ and $K$&$C_s$& $A'$ & $A'$ & $A'$ & $A'_1$, $E'$ \\
    \hline
    \end{tabular}
    \caption{Group theory analysis for exciton-phonon interactions for excitons A and C of Mo$_2$}
    \label{tab:group_theory_analysis}
\end{table*}

\begin{figure}
    \centering
    \includegraphics[width=\linewidth]{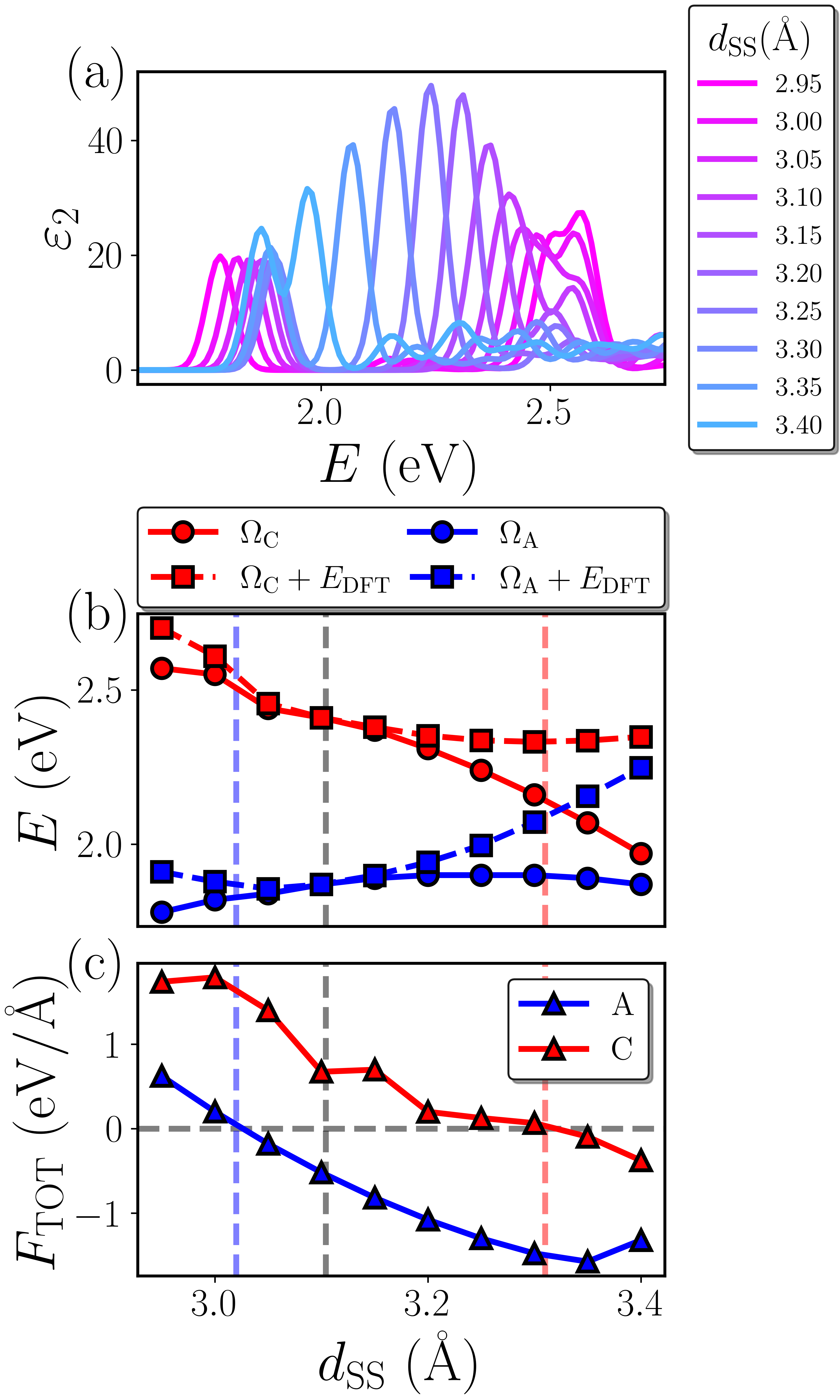}
    \caption{Relaxation of excited states in 1L MoS2. (a) Optical absorption of 1L MoS2 by changing the out-of-plane distance $d_{\rm{SS}}$ between sulfur atoms ($A'_1$ phonon mode). As $d_{\rm{SS}}$ increases, the A peak is blueshifted while the C peak is redshifted. (b) Solid lines are exciton energies and dashed lines total energy (exciton and DFT energies summed) of the brightest excitons in A (in blue) and C (in red) peaks. (d) Total force due to excitons of panel (b). The vertical black line shows the equilibrium $d_{\rm{SS}}$ distance for the ground state, while blue and red lines show the equilibrium $d_{\rm{SS}}$ for the total energy curves of excitons A and C, respectively.}
    \label{fig:1lmos2_exc_energy_vs_dSS}
\end{figure}

As the $A'_1$ phonon mode presents the strongest exciton-phonon coupling we studied how its displacement affects A and C excitons in monolayer MoS$_2$. We found that when the distance between sulfur atoms ($d_{\rm{SS}}$ in Figure \ref{fig:1lmos2_exc_energy_vs_dSS}) is increased C (A) exciton energy is redshifted (blueshifted), as shown in panel (a) of Figure \ref{fig:1lmos2_exc_energy_vs_dSS}. In panel (b) we plot the total energy for the extreme case of one exciton per primitive cell ($\sim 1.2 \times 10^{15}$ exciton / cm$^2$). The trapping energy for the exciton in the C peak is 78 meV with increase of the $d_{\rm{SS}}$ by 0.2 \text{\AA}, while for exciton A the trapping energy is 14 meV and $d_{\rm{SS}}$ decreases by 0.08 F\text{\AA}. Experimentally, typical exciton concentrations are up to 10$^{12}$ excitons / cm$^2$, leading to exciton trapping energies smaller than 1 meV and distortion of $d_{\rm{SS}}$ are around $10^{-4}$ \text{\AA}, which is not enough to be detectable by AFM experiments but strong enough to generate resonant phonons \cite{Trovatello2020}. The energy curves in figure \ref{fig:1lmos2_exc_energy_vs_dSS}, are useful in understanding the mechanism of excitation of coherent phonons by light absorption \cite{Trovatello2020, Zeiger1992}.

\section{ESF and Spin Orbit Coupling}

\label{sec:ESF_and_SOC}

Now we discuss the ESF when including Spin Orbit Coupling (SOC). In the case of semiconductors where the SOC splits are much smaller than the material bandgap, one can include the SOC interactions perturbatively as it is done in the references \cite{Qiu2016, QiuPRL2013}. In this approximation the exciton energy is given by $\Omega^A_\sigma = \Omega^A + \Delta \Omega^A_\sigma$, where the second term is the perturbation correction given by $\Delta \Omega^A_\sigma = \sum_{\mathbf{k}cv} |A_{\mathbf{k}cv}|^2 \left[ (E^{\rm{QP}}_{\mathbf{k}c} + \Delta E^{\rm{QP}}_{\mathbf{k}c\sigma}) - (E^{\rm{QP}}_{\mathbf{k}v} + \Delta E^{\rm{QP}}_{\mathbf{k}v\sigma}) \right]$, where $\Delta E^{\rm{QP}}_{\mathbf{k}i\sigma} = E^{\rm{QP-FR}}_{\mathbf{k}i\sigma} - E^{\rm{QP-SR}}_{\mathbf{k}i} \approx E^{\rm{DFT-FR}}_{\mathbf{k}i\sigma} - E^{\rm{DFT-SR}}_{\mathbf{k}i} $ is the variation of GW calculations including SOC effects in relation to GW calculation without SOC effects, and the QP difference is approximated to be equal to the DFT difference. This approximation is justified in case of small SOC effects, as for example in MoS$_2$, where the valence SOC split in the $K$ point is close to the experimental difference between A and B excitons \cite{Qiu2016, QiuPRL2013}. With this approximation, the excited state forces have a correction given by

\begin{equation}
    \Delta F = - \nabla \Delta \Omega^A_\sigma = 
    -\sum_{\mathbf{k} cv} |A_{\mathbf{k}cv}|^2 
    \left(
        \nabla\Delta E^{\rm{QP}}_{\mathbf{k}c\sigma}
        - \nabla\Delta E^{\rm{QP}}_{\mathbf{k}v\sigma}
    \right)
\end{equation}

\begin{figure}
    \centering
    \includegraphics[width=\linewidth]{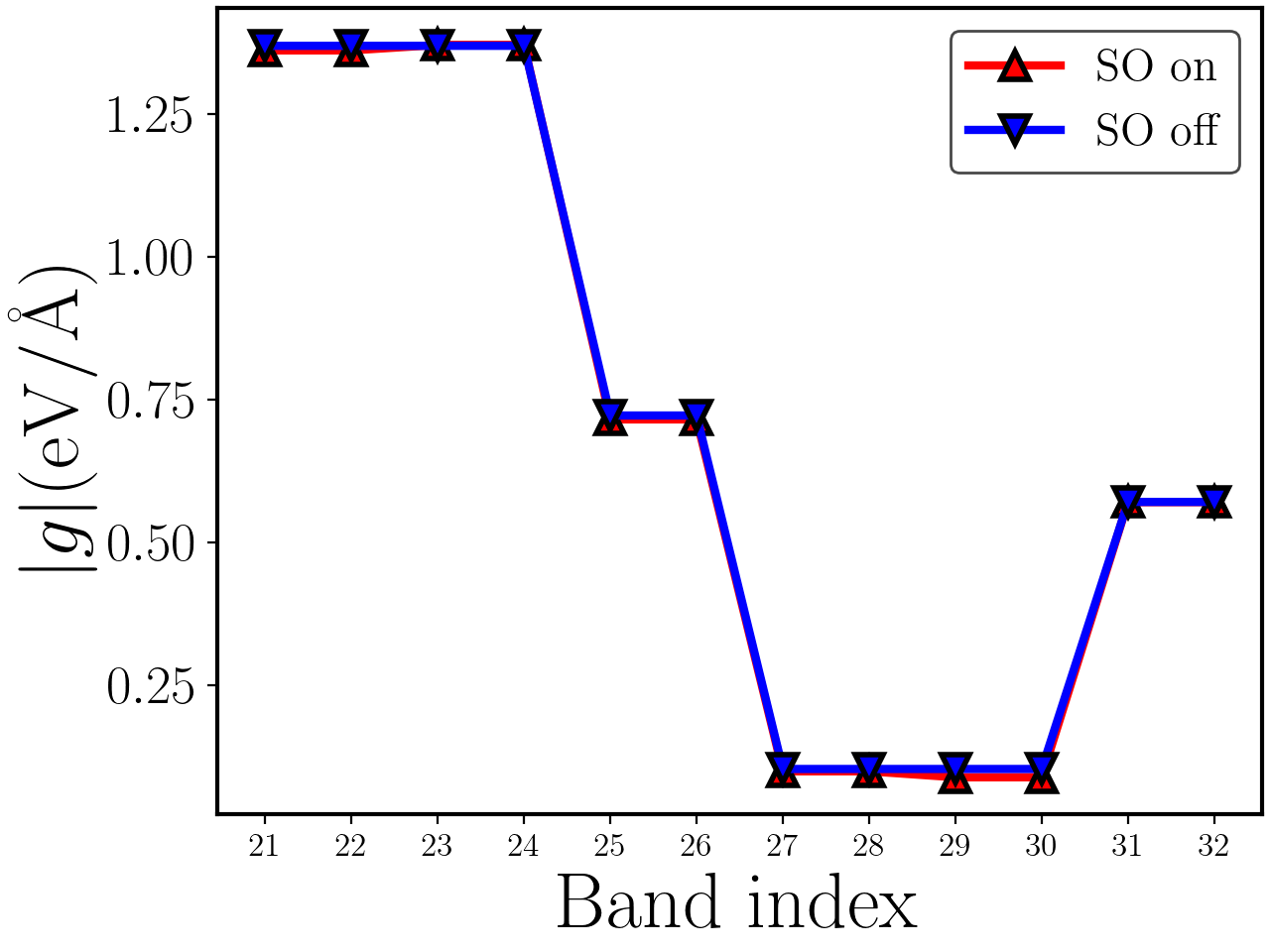}
    \caption{Diagonal electron-phonon coefficients for monolayer MoS$_2$ with and without SOC. The mode shown here is the $A'_1$, where the two sulfur atoms move towards each other and the molybdenum atom does not move.}
    \label{fig:elph_coeffs_MoS2_SOC_influence}
\end{figure}

In the above equation, the gradient $\nabla\Delta E^{\rm{QP}}_{\mathbf{k}i\sigma}$ is equal to the difference of diagonal ELPH coefficients calculated with and without SOC interactions. To understand the importance of this term we analyze the difference between those two terms for the $A'_1$ phonon mode in MoS$_2$ monolayer, where the two sulfur atoms move perpendicular to the monolayer plane in opposite directions and the molybdenum atom remains stopped. We observe in figure \ref{fig:elph_coeffs_MoS2_SOC_influence} that this difference is negligible in comparison to the magnitude of the diagonal ELPH coefficients, therefore the ESF theory can be safely applied in systems where SOC can be treated with perturbation theory.

\section{Conclusions} 
\label{sec:conclusions}

We have presented and implemented a practical workflow to calculate ESF from GW-BSE calculations that combines excitonic effects with electron-phonon coefficients from DFPT, based on reference \cite{IsmailBeigi2003}. We solved the non-zero force on the system center of mass issue and studied extensively the quality and validity of approximations used in our approach. We applied a renormalization scheme for electron-phonon coefficients, which is simple and that improves the agreement of our excited state forces calculations with finite differences. The ESF calculations, in general takes only few minutes in personal computers and much more faster than GWBSE and DFPT calculations. Those forces can be projected into phonon displacement basis to study exciton-phonon interactions \cite{Karkee2026AdvMatTech}. We performed relaxations of self-trapped polaronic excitons in LiF, in agreement with previous theoretical works \cite{Dai2024PRB, Dai2024PRL}. In reference \cite{DelGrande2026arxiv}, we found that most excitons lead to repulsive interactions between layers in WSe$_2$, in agreement with the experimental observation of light-induced 0.2 \text{\AA} increase in the interlayer distance of 0.3$^\circ$ twisted bilayer WSe$_2$\ \cite{NakamuraAdvMat2026}. The excited state forces are diagonal exciton-phonon matrix elements ($\partial_\nu \Omega_A =\langle A|\partial_\nu H^{\rm{BSE}} |A\rangle$), but our code is also capable of computing off-diagonal exciton-phonon matrix elements with transfer of momentum $\langle A(Q)| \partial_{\nu, q} H^{\rm{BSE}}|B(Q+q)\rangle$, which can be used within computational studies of resonant Raman, photoluminescence, and time evolution of exciton populations, and that functionalities will be discussed in future works. Those forces can also be used to train machine learning force field models \cite{Westermayr2020} which would speed up the study of the dynamics of excited states. 
Therefore the ESF theory can be applied to the understanding of microscopic mechanisms of light-matter interaction.

\section*{Acknowledgments}

R.R.D.G. and D.A.S. were supported by the U.S. National Science Foundation under Grant No. DMR-2144317 and Cottrell Scholar award No. 26921, a program of Research Corporation for Science Advancement; and this material is based upon work supported by the Air Force Office of Scientific Research under award number FA9550-19-1-0236.
Computational resources were provided by the National Energy Research Scientific Computing Center (NERSC), a U.S. Department of Energy Office of Science User Facility operated under Contract No. DE-AC02-05CH11231; the Texas Advanced Computing Center (TACC) at The University of Texas at Austin (http://www.tacc.utexas.edu); and the Pinnacles and Multi-Environment Computer for Exploration and Discovery (MERCED) clusters at UC Merced, funded by National Science Foundation Grants No. OAC-2019144 and ACI-1429783.


\onecolumngrid

\section{Appendix}

\subsection{Excited state forces equations}
\label{sec:ESF_equations}

Here we dedicate a section discussing the calculation of the gradient of the excitation energy with two different approaches: our approach where we make use of the Hellman-Feynman theorem and  later change its basis to $|\mathbf{k}cv\rangle$ and the approach used by \cite{IsmailBeigi2003} where first the basis is changed to $|\mathbf{k}cv\rangle $ and later the Hellman-Feynman theorem is used. The BSE Hamiltonian is given by $H^{\rm{BSE}} = D +K^{\rm{eh}}$, where $D = H^{\rm{QP}}_c \otimes I_v - I_c \otimes H^{\rm{QP}}_v$ is the independent particle transition part, composed by the single particle GW Hamiltonian $H^{\rm{QP}}_{c(v)} |\mathbf{k}c(v)\rangle = E^{\rm{QP}}_{\mathbf{k}c(v)} |\mathbf{k}c(v)\rangle$, and $K^{\rm{eh}}$ is the electron-hole kernel (see text). The excitation energy $\Omega_A$ is eigenvalue of the BSE Hamiltonian $H^{\rm{BSE}}|A\rangle=\Omega_A |A\rangle$, where $|A\rangle$ is the exciton state that can be expanded in the basis of transitions from valence states $|\mathbf{k}v \rangle$ to conduction states $|\mathbf{k}c\rangle$, represented by $|\mathbf{k}cv\rangle = |\mathbf{k}c\rangle \otimes |\mathbf{k}v\rangle$, in the form $|A\rangle = \sum_{\mathbf{k}cv} A_{\mathbf{k}cv} |\mathbf{k}cv\rangle$, where $A_{\mathbf{k}cv} = \langle \mathbf{k}cv | A\rangle$. The independent particle transition part in the $|\mathbf{k}cv\rangle$ basis is given by $\langle \mathbf{k}cv |D|\mathbf{k}'c'v'\rangle=\delta_{\mathbf{kk}'} \delta_{cc'} \delta_{vv'} (E^{\rm{QP}}_{\mathbf{k}c}-E^{\rm{QP}}_{\mathbf{k}v})$. 

\subsubsection{Differentiate before change basis}

First invoking the Hellman-Feynman theorem we have that

\begin{equation}
    \partial \Omega_A = \langle A | \partial D + \partial K^{\rm{eh}} |A \rangle 
\end{equation}

and we can expand the coefficients $A$ in the $\mathbf{k}cv$ basis 

\begin{equation}
\begin{split}
    \partial \Omega_A = & \sum_{\mathbf{k} cv \mathbf{k}'c'v'} \langle A | \mathbf{k} cv \rangle \langle \mathbf{k} cv| \partial D + \partial K^{\rm{eh} } | \mathbf{k}'c'v' \rangle \langle \mathbf{k}'c'v'|A\rangle \\
    & \sum_{\mathbf{k} cv \mathbf{k}'c'v'} A_{\mathbf{k} cv} A^*_{\mathbf{k}'c'v'} \langle \mathbf{k}cv |\partial D + \partial K^{\rm{eh}} |\mathbf{k}'c'v'\rangle 
\end{split}
\end{equation}

As the independent particle transition part is given by $D=H^{\rm{QP}}_c \otimes I_v - I_c \otimes H^{\rm{QP}}_v $, its derivative is $\partial D = \partial H^{\rm{QP}} \otimes I_v - I_c \otimes \partial H^{\rm{QP}}_v $, then 
\begin{equation}
    \langle \mathbf{k}cv | \partial D | \mathbf{k}'c'v' \rangle = (\langle \mathbf{k}c | \partial H^{\rm{QP}} | \mathbf{k}'c' \rangle \delta_{vv'} - \langle \mathbf{k}v |\partial H^{\rm{QP}} | \mathbf{k}v'\rangle \delta_{cc'} )\delta_{\mathbf{k} \mathbf{k}'}
    \label{eq:derivative_D_our_method}
\end{equation}
Therefore the excited state force expression becomes
\begin{equation}
    \partial \Omega_A = \sum_{\mathbf{k} cv \mathbf{k}'c'v'} A_{\mathbf{k} cv} A^*_{\mathbf{k}'c'v'} 
    \left( 
    (\langle \mathbf{k}c | \partial H^{\rm{QP}} | \mathbf{k}'c' \rangle \delta_{vv'} - \langle \mathbf{k}v |\partial H^{\rm{QP}} | \mathbf{k}v'\rangle \delta_{cc'} )\delta_{\mathbf{k} \mathbf{k}'}  
    + \langle \mathbf{k}cv | \partial K^{\rm{eh}} | \mathbf{k}'c'v'\rangle 
    \right)
    \label{eq:ESF_our_equation_derived}
\end{equation}
which is our expression for ESF with the approximation of $\langle \mathbf{k}cv | \partial K^{\rm{eh}} | \mathbf{k}'c'v'\rangle \approx 0$. 

\subsubsection{Change basis before differentiate}

 First we rewrite $\Omega_A = \langle A | D + K^{\rm{eh}} | A\rangle $ expanded in the $|\mathbf{k}cv\rangle$ basis
 \begin{equation}
     \Omega_A = \sum_{\mathbf{k} cv \mathbf{k}'c'v'} \langle A|\mathbf{k}cv\rangle \langle \mathbf{k}'c'v' | A \rangle \langle \mathbf{k}cv | D +K^{\rm{eh}} |\mathbf{k}'c'v'\rangle 
 \end{equation}
Then its derivative will be 
\begin{equation}
\begin{split}
\partial \Omega_A  = \sum_{\mathbf{k} cv \mathbf{k}'c'v'}  &
\partial \langle A|\mathbf{k}cv\rangle \langle \mathbf{k}'c'v' | A \rangle \langle \mathbf{k}cv | D +K^{\rm{eh}} |\mathbf{k}'c'v'\rangle   \\
& + \langle A|\mathbf{k}cv\rangle \partial \langle \mathbf{k}'c'v' | A \rangle \langle \mathbf{k}cv | D +K^{\rm{eh}} |\mathbf{k}'c'v'\rangle   \\
& + \langle A|\mathbf{k}cv\rangle \langle \mathbf{k}'c'v' | A \rangle \partial \langle \mathbf{k}cv | D +K^{\rm{eh}} |\mathbf{k}'c'v'\rangle 
\end{split}
\end{equation}

The first term is given by

\begin{equation}
\begin{split}
\sum_{\mathbf{k} cv \mathbf{k}'c'v'} 
& \partial \langle A|\mathbf{k}cv\rangle \langle \mathbf{k}'c'v' | A \rangle \langle \mathbf{k}cv | H^{\rm{BSE}} |\mathbf{k}'c'v'\rangle \\
= & \sum_{\mathbf{k}cv} \partial \langle A|\mathbf{k}cv\rangle
\sum_{\mathbf{k}'c'v'}  \langle \mathbf{k}'c'v' | A \rangle \langle \mathbf{k}cv | H^{\rm{BSE}} |\mathbf{k}'c'v'\rangle  \\
= &  \Omega_A \sum_{\mathbf{k}cv} \partial \langle A|\mathbf{k}cv\rangle  \langle \mathbf {k} cv | A \rangle 
\end{split}
\end{equation}
Where we used the identity $\sum_{\mathbf{k}cv} \langle A| \mathbf{k}cv\rangle \langle \mathbf{k}cv | H^{\rm{BSE}} |\mathbf{k}'c'v'\rangle = \Omega_A \langle A| \mathbf{k}'c'v'\rangle $. Similarly we have that second term is given by
\begin{equation}
\begin{split}
\sum_{\mathbf{k} cv \mathbf{k}'c'v'} 
&  \langle A|\mathbf{k}cv\rangle \partial \langle \mathbf{k}'c'v' | A \rangle \langle \mathbf{k}cv | H^{\rm{BSE}} |\mathbf{k}'c'v'\rangle = 
\Omega_A \sum_{\mathbf{k}'c'v'} \langle A|\mathbf{k}'c'v'\rangle  \partial \langle \mathbf{k}'c'v' | A\rangle 
\end{split}
\end{equation}
Putting those two terms together we have that 
\begin{equation}
\begin{split}
    & \Omega_A \sum_{\mathbf{k}cv} \partial \langle A|\mathbf{k}cv\rangle  \langle \mathbf {k} cv | A \rangle +  \langle A|\mathbf{k}cv\rangle \partial \langle \mathbf {k} cv | A \rangle \\
    = & \Omega_A \partial \left( \sum_{\mathbf{k}cv} \langle A|\mathbf{k}cv\rangle \langle \mathbf {k} cv | A \rangle \right) = \Omega_A \partial (\langle A|A\rangle ) = \Omega_A \partial (1) = 0
\end{split}
\end{equation}
Therefore, due to this cancellation we have that 
\begin{equation}
    \partial \Omega_A = \sum_{\mathbf{k} cv \mathbf{k}'c'v'} A_{\mathbf{k} cv} A^*_{\mathbf{k}'c'v'} \partial \langle \mathbf{k}cv | D+K^{\rm{eh}}|\mathbf{k}'c'v'\rangle 
\end{equation}
which is the equation used by Ismail-Beigi and Louie \cite{IsmailBeigi2003}. In other words, this is the Hellman-Feynman theorem, but by applying derivatives after a change of basis from $|A\rangle$ to $|\mathbf{k}cv \rangle$. Now, first we calculate the derivative including $D$. In reference \cite{IsmailBeigi2003} the independent particle part is given by $D = (E^{\rm{QP}}_{\mathbf{k}c} - E^{\rm{QP}}_{\mathbf{k}v})\delta_{\mathbf{kk}'} \delta_{vv'} \delta_{cc'}$. A derivation using the definition of $D=H^{\rm{QP}}_c \otimes I_v - I_c \otimes H^{\rm{QP}}_v$ we have that
\begin{equation}
    \partial \langle \mathbf{k}cv | D|\mathbf{k}'c'v' \rangle = \partial (\langle \mathbf{k}c | H^{\rm{QP}}_c | \mathbf{k}c'\rangle \delta_{vv'} -
    \langle \mathbf{k}v | H^{\rm{QP}}_v | \mathbf{k}v'\rangle \delta_{cc'}) \delta_{\mathbf{k} \mathbf{k}'}
\label{eq:ibl_derivative_matrix_element_D}
\end{equation}
and for that we need to calculate the derivative of the matrix elements $\langle \mathbf{k}i|  H^{\rm{QP}} | \mathbf{k}j \rangle$, where $|\mathbf{k}i\rangle$ is eigenvector of $H^{\rm{QP}}$. In general, a derivative of a matrix element of an one-body Hamiltonian $H$ with eigenstates $|i\rangle$ and for $i\neq j$ is given by
\begin{equation}
\begin{split}                     
    \partial \langle i | H | j \rangle  = &   
    \langle  \partial i | H | j\rangle  
    +   \langle i | H | \partial j\rangle 
    +   \langle i |\partial  H | j\rangle \\
= & E_j \langle \partial i | j \rangle + E_i \langle i | \partial  j\rangle + \langle i | \partial H | j \rangle
\end{split}
\label{eq:one-body-hamiltonian-offdiag-derivative}
\end{equation}
Expanding the derivatives of the single particle states with first order perturbation theory we get that
\begin{equation}
    \langle i | \partial j \rangle = \langle i | \sum_{l\neq j} \frac{\langle l | \partial H |j \rangle}{E_j -E_l} |l\rangle = \frac{\langle i | \partial H|j\rangle}{E_j - E_i}
\end{equation}
and by orthogonality $\langle \partial i |  j \rangle = -\langle i | \partial j\rangle$. So equation \ref{eq:one-body-hamiltonian-offdiag-derivative} becomes
\begin{equation}
\begin{split}
    \partial \langle i | H | j\rangle = & -E_j \frac{\langle i | \partial H | j \rangle}{E_j - E_i}  + E_i  \frac{\langle i | \partial H | j \rangle}{E_j - E_i}  + \langle i | \partial H|j \rangle\\
    = & -(E_j  - E_i ) \frac{\langle i | \partial H | j \rangle  }{E_j - E_i} + \langle i | \partial H|j \rangle  = 0
\end{split}
\end{equation}

While for the case of $i = j$ then we can directly use Hellman-Feynman theorem and then $\partial \langle i | H | j \rangle = \partial  E_i \delta_{ij}$. Therefore equation \ref{eq:ibl_derivative_matrix_element_D} becomes
\begin{equation}
    \partial \langle \mathbf{k}cv | D|\mathbf{k}'c'v' \rangle = (\partial E^{\rm{QP}}_{\mathbf{k}c} - \partial E^{\rm{QP}}_{\mathbf{k}v})\delta_{cc'} \delta_{vv'} \delta_{\mathbf{k}\mathbf{k'}}
xt    \label{eq:ibl_rpa_part_just_diag}
\end{equation}
Notice that $\partial \langle \mathbf{k}cv | D|\mathbf{k}'c'v' \rangle$ does not include any off-diagonal ELPH matrix element, differently from equation \ref{eq:derivative_D_our_method}. 

For the kernel derivatives the derivative is given by
\begin{equation}
\begin{split}
\partial \langle \mathbf{k}cv | K^{\rm{eh}} | \mathbf{k}'c'v' \rangle  = 
    & 
    \sum_{c'' \neq c} \langle \mathbf{k}c''v|K^{\rm{eh}}|\mathbf{k}c'v'\rangle \frac{\langle \mathbf{k}c| \partial H^{\rm{QP}} | \mathbf{k}c'' \rangle}{E^{\rm{QP}}_{\mathbf{k}c} - E^{\rm{QP}}_{\mathbf{k}c''}} \\ 
+ & \sum_{c'' \neq c'} \langle \mathbf{k}cv | K^{\rm{eh}} | \mathbf{k}'c''v'\rangle \frac{\langle \mathbf{k}'c''| \partial H^{\rm{QP}} | \mathbf{k}'c' \rangle}{E^{\rm{QP}}_{\mathbf{k}'c'} - E^{\rm{QP}}_{\mathbf{k}'c''}} \\
 + & \sum_{v'' \neq v} \langle \mathbf{k}cv'' | K^{\rm{eh}} | \mathbf{k}'c'v'\rangle \frac{\langle \mathbf{k}v| \partial H^{\rm{QP}} | \mathbf{k}v'' \rangle}{E^{\rm{QP}}_{\mathbf{k}v} - E^{\rm{QP}}_{\mathbf{k}v''}} \\
  + & \sum_{v'' \neq v'} \langle \mathbf{k}cv | K^{\rm{eh}} | \mathbf{k}'c'v''\rangle \frac{\langle \mathbf{k}'v''| \partial H^{\rm{QP}} | \mathbf{k}'v' \rangle}{E^{\rm{QP}}_{\mathbf{k}'v'} - E^{\rm{QP}}_{\mathbf{k}'v''}} \\
 + & \langle  \mathbf{k}cv | \partial K^{\rm{eh}} | \mathbf{k}'c'v' \rangle
\end{split}
\label{eq:expansion_kernel_derivatives_ibl}
\end{equation}
where the last term is approximated to be zero. In reference \cite{IsmailBeigi2003}, the above equation is also used with the approximation $\langle \mathbf{k}i| \partial H^{\rm{QP}}|\mathbf{k}j \rangle \approx \langle \mathbf{k}i| \partial H^{\rm{DFT}}|\mathbf{k}j \rangle$, while in this work we evaluate the above equation using our renormalization scheme from equation \ref{eq:renormalization_elph_coeffs}. The ESF in reference \cite{IsmailBeigi2003} is the sum of equations \ref{eq:ibl_rpa_part_just_diag} and \ref{eq:expansion_kernel_derivatives_ibl}. By comparing those with the main equation used in this work (equation \ref{eq:ESF_our_equation_derived}) one can notice that the first four terms in equation \ref{eq:expansion_kernel_derivatives_ibl} are equal to the off-diagonal part of equation \ref{eq:derivative_D_our_method}. 

\twocolumngrid
\subsection{Renormalization of ELPH coefficients }

\label{sec:renormalization_ELPH_coefficients_equations}
We start from the definition of the QP hamiltonian matrix elements

\begin{equation}
    H^{\rm{QP}}_{ij} = H^{dft}_{ij} \delta_{ij} + \Delta \Sigma_{ij}\delta_{ij} 
\end{equation}

\noindent where $|i\rangle$ is an eigenstate of the DFT hamiltonian with eigenvalue $E^{dft}_i$ and $\Delta \Sigma_{ij} = \langle i | \Sigma - V_{xc} | j \rangle$.  The state $|i\rangle$ is also eigenstate of the QP hamiltonian. Using, first order perturbation theory we have that 
\begin{equation}
    |\partial i \rangle = \sum_{k \neq i} \frac{\langle k | \partial H^{dft} | i \rangle}{E^{dft}_i - E^{dft}_k} |k\rangle
\end{equation}

But as this state is also an eigenstate of the QP hamiltonian, we can write

\begin{equation}
    |\partial i \rangle = \sum_{k \neq i} \frac{\langle k | \partial H^{\rm{QP}} | i \rangle}{E^{qp}_i - E^{qp}_k} |k\rangle
\end{equation}
This implies that 

\begin{equation}
    \frac{\langle j | \partial H^{dft} | i \rangle}{E^{dft}_i - E^{dft}_j} = 
    \frac{\langle j | \partial H^{\rm{QP}} | i \rangle}{E^{qp}_i - E^{qp}_j} 
\end{equation}

\noindent which leads to equation \ref{eq:renormalization_elph_coeffs}. 


\subsection{Computational Details}
\label{sec:computational_details}

We used ONCV scalar-relativistic LDA pseudopotentials with standard precision from Pseudodojo \cite{VansettenCompPhysCom2018, HamannPRB2013}. All G$_0$W$_0$ calculations were performed with Generalized Plasmon Pole approximation for frequency dependency of the self-energy \cite{Hybertsen1986} and the static remainder coulomb-hole \cite{DeslippePRB2013}. 

\subsubsection{CO}
For CO DFT calculations the kinetic energy cutoff is equal to 100 Ry and the k-grid was composed just by the $\Gamma$ point. In GW calculations the dielectric matrix was built with 500 empty states and a cutoff of 20 Ry for the screened Coulomb interaction and truncate the Coulomb interaction to avoid interactions with its periodic images \cite{IBPRB2006}. The BSE hamiltonian was built with 5 valence states and 13 conduction states, except when stated differently. 

\subsubsection{LiF}
For primitive cell LiF DFT calculaitons the kinetic energy cutoff is equal to 80 Ry and the k-grid was a regular k-grid $4\times 4 \times 4$ centered at the $\Gamma$ point. The relaxed lattice parameter was 2.0295 $\text{\AA}$. For GW/BSE calculations we used a coarse grid of $4\times 4 \times 4$ and a fine grid of $8\times 8 \times 8$, both centered at $\Gamma$. In GW calculations the dielectric matrix was built with 300 empty states and a cutoff of 20 Ry for the screened Coulomb interaction. The BSE hamiltonian was built with 5 valence states and 10 conduction states. For GW/BSE calculations in 4$\times$4$\times$4 supercell we used the stochastic pseudobands method \cite{AltmanPRL2024} with an accumulation window of 2\%, 2 stochastic pseudobands per energy subspace, and generating empty states with energy up to the DFT cutoff (80 Ry). We build the BSE Hamiltonian with 31 conduction and 137 valence bands based on our approach to make appropriate choices of the BSE Hamiltonian size, where we reduced the number of kernel matrix elements to be calculated to 12\% from a zone folding analisys with an error of 0.15 eV for the exciton energy \cite{delgrande2025}. 

\subsubsection{\texorpdfstring{MoS$_2$}{MoS2}}
For monolayer MoS$_2$ DFT calculations were performed with a cutoff energy of 80 Ry, with in plane and out of plane lattice parameters of 3.12 \text{\AA} and 25 \text{\AA}, respectively. We truncate the Coulomb interaction in the direction perpendicular to sheet layer \cite{SohierPRB2017}. We performed G$_0$W$_0$ calculations within the Generalized Plasmon Pole, with a cutoff energy to build the dielectric matrix equal to 25 Ry, and the stochastic pseudobands method \cite{AltmanPRL2024} with an accumulation window of 2\%, 2 stochastic pseudobands per energy subspace, and generating empty states with energy up to the DFT cutoff (80 Ry). Also for GWBSE calculations we use a Coulomb truncation for 2D systems \cite{IBPRB2006}. Quasi particle (QP) energies and kernel matrix elements were calculated in a coarse $k$-grid 6$\times$6$\times$1 and interpolated into a fine $k$-grid 18$\times$18$\times$1 to build the BSE Hamiltonian. To improve the convergence with respect to $k$-grid in both GW and BSE calculations, we used non-uniform sampling methods \cite{JornadaPRB2017}.

\subsection{\texorpdfstring{Tight binding model of ESF on H$^+_2$}{Tight binding model of ESF on H2+}}
\label{sec:tight_binding_model_H2}

To understand the physics of the excited state forces, we will analyze the H$_2^+$ molecule. In a tight binding model, the HOMO and LUMO energy levels are given by \cite{atkins2011Book}

\begin{equation}
    E_{\rm{HOMO(LUMO)}} = E_{\rm{1s}} + \frac{j_0}{R} + \frac{j \pm k}{1 \mp S}
\end{equation}
\noindent where $j_0 = e^2 / 4 \pi \epsilon_0$ ($e$ is the electron charge and $\epsilon_0$ the vacuum dielectric constant), $E_{\rm{1s}}$ = 13.6 eV, $R$ is the distance between the hydrogen atoms, $a_0$ is the Bohr radius, the other quantities are given by

\begin{equation}
    j = \frac{j_0}{R} \left( 1 - \left( 1 + \frac{R}{a_0}  \right) e^{-2R/a_0}  \right)
\end{equation}
\begin{equation}
    k = \frac{j_0}{a_0} \left( 1 + \frac{R}{a_0} \right) e^{-R/a_0}
\end{equation}
\noindent and the overlap $S$ is given by

\begin{equation}
    S = \left( 1 + \frac{R}{a_0} + \frac{1}{3} \frac{R^2}{a_0^2} \right) e^{-R/a_0}
\end{equation}
The bandgap is given by

\begin{equation}
    E_g = \frac{2(k-Sj)}{1-S^2}
\end{equation}
An exciton composed of the HOMO and LUMO states has energy equal to $\Omega = E_g + K$. In this case, we approximate the kernel derivative to be zero, which is a common approximation we use in the main text. An analytical expression for the derivative of the gap is quite complicated. We can analyze the plot of $E_g$ as a function of the nucleus distance. We observe that the gap decreases when $R$ increases, as the difference between bonding and antibonding states becomes smaller. Therefore when an exciton is composed of a pair of bonding and antibonding states, which is common in simple molecules, the ESF will be repulsive.

\begin{figure}[ht]
\includegraphics[width=\columnwidth]{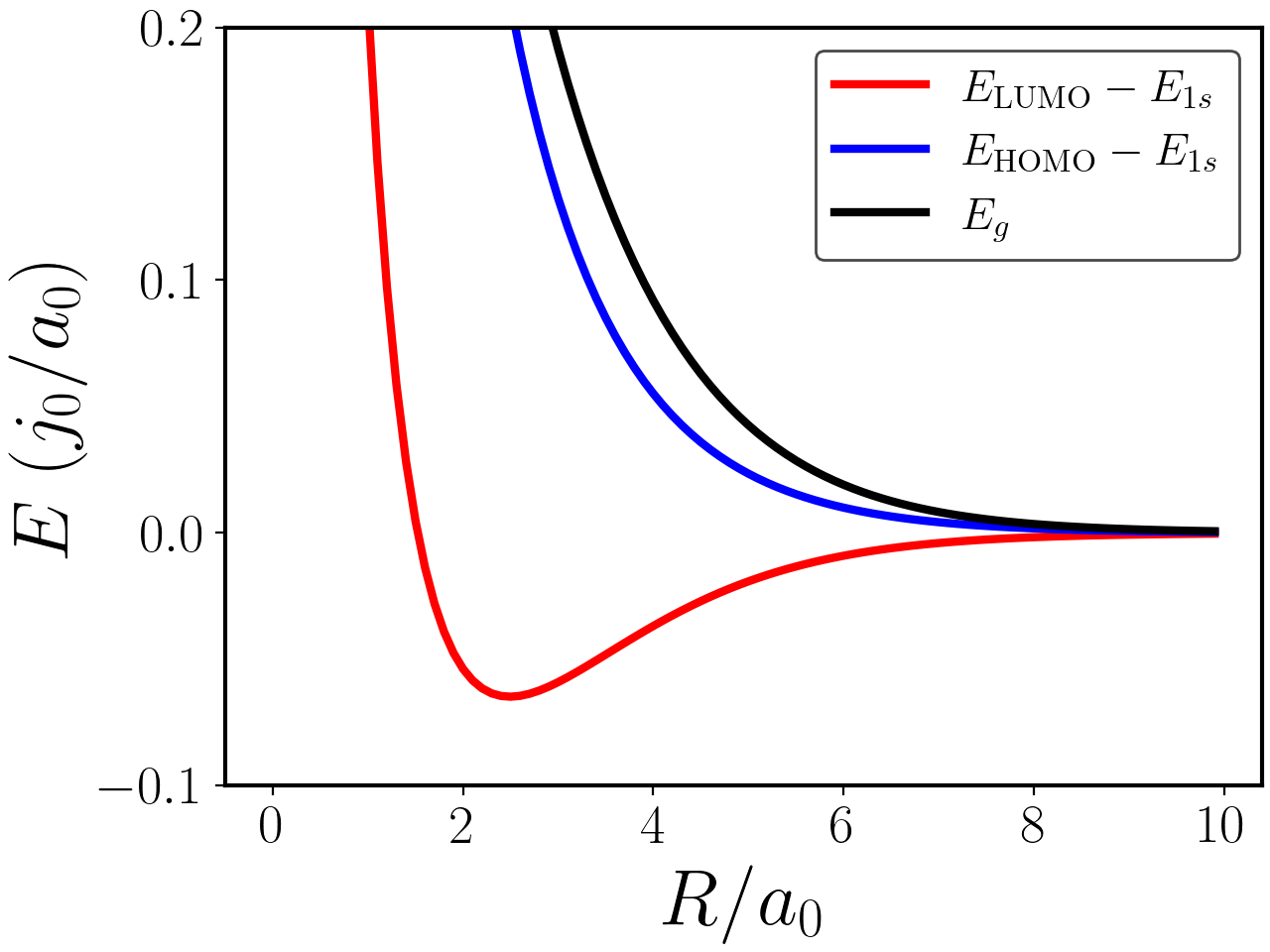}
\caption{HOMO, LUMO and gap energies in a H$_2^+$ molecule using a tight binding model. }
\centering
\label{fig:H2_model}
\end{figure}

\bibliography{refs}

\end{document}